\documentclass[sn-mathphys,Numbered]{sn-jnl}

\usepackage{graphicx}%
\usepackage{multirow}%
\usepackage{amsmath,amssymb,amsfonts}%
\usepackage{amsthm}%
\usepackage{mathrsfs}%
\usepackage[title]{appendix}%
\usepackage{xcolor}%
\usepackage{textcomp}%
\usepackage{manyfoot}%
\usepackage{booktabs}%
\usepackage{listings}
\usepackage{graphicx}
\usepackage{dcolumn}
\usepackage{bm}
\usepackage{subcaption}

\usepackage{xcolor}
\usepackage{physics}
\usepackage{float}
\usepackage{algorithmic}  
\usepackage[algo2e]{algorithm2e} 
\usepackage[]{algorithm2e}
\usepackage{algorithm}

\usepackage{url}
\usepackage[english]{babel}
\usepackage{amsthm, amssymb}
\usepackage{cleveref}
\usepackage{enumitem}
\usepackage{caption}
\DeclareCaptionLabelSeparator{period}{.\quad}
\theoremstyle{definition}

\theoremstyle{definition}

\DeclareCaptionLabelFormat{mylabel}{ FIG. #2} 
\DeclareCaptionLabelFormat{mytabel}{ TABLE. #2} 

\begin{document}

\title{NISQ-friendly measurement-based quantum clustering algorithms}

\author[1,2]{\fnm{Srushti} \sur{Patil}}\email{srushtipatil926@gmail.com}

\author*[3,4]{\fnm{Shreya} \sur{Banerjee}}\email{shreya93ban@gmail.com}

\author[3,5]{\fnm{Prasanta K.} \sur{Panigrahi}}\email{director.cqst@soa.ac.in}

\affil[1]{\orgdiv{\emph{NNF Quantum Computing Programme, Neils Bphr Institute}}, \orgname{University of Copenhagen} \orgaddress{ \state{Copenhagen N DK-2200}, \country{Denmark}}}

\affil*[2]{\orgdiv{Department of Physics}, \orgname{Indian Institute of Science Education and Research Tirupati}, \orgaddress{\state{Andhra Pradesh, 517619}, \country{India}}}

\affil[3]{\orgdiv{Center for Quantum Science and Technology}, \orgname{Siksha 'O' Anusandhan University} \orgaddress{ \state{Bhubaneswar, Odisha 751030}, \country{India}}}

\affil[4]{\orgname{Institut quantique de l'Universit\'e de Sherbrooke}, \orgaddress{ \state{Queb\'ec, J1K2R1} \country{Canada}}}

\affil[5]{\orgname{Department of Physical Sciences, Indian Institute of Science Education and Research Kolkata}, \orgaddress{ \state{West Bengal, 741246}, \country{India}}}

\maketitle

\begin{abstract}
{Two novel measurement-based, quantum clustering algorithms are proposed based on quantum parallelism and entanglement. The first algorithm follows a divisive approach. The second algorithm is based on unsharp measurements, where we construct an effect operator with a Gaussian probability distribution to cluster similar data points. A major advantage of both algorithms is that they are simplistic in nature, easy to implement, and well suited for noisy intermediate scale quantum computers. We have successfully applied the first algorithm on a concentric circle data set, where the classical clustering approach fails, as well as on the Churrtiz data set of $130$ cities, where we show that the algorithm succeeds with very low quantum resources. We applied the second algorithm on the labeled Wisconsin breast cancer dataset, and found that it is able to classify the dataset with high accuracy using only $O(log(D))$ qubits and polynomial measurements, where $D$ is the maximal distance within any two points in the dataset. We also show that this algorithm works better with an assumed measurement error in the quantum system, making it extremely well-suited for NISQ devices.}
\end{abstract}

\keywords{Quantum Clustering, Unsharp Measurements, Quantum Algorithms}
\section{Introduction}\label{sec 1}
 Quantum supremacy over classical computers is currently a topical subject of investigation \cite{arute2019quantum}, where demonstrating quantum advantage is particularly challenging. The performance of certain classical algorithms can be improved with the help of quantum computers. To achieve speedup, quantum algorithms are being developed inspired by the fundamental quantum laws like entanglement and the superposition principle leading to quantum parallelism. The potential of quantum computation was first realized by Feynman in 1982, where he postulated the need for quantum computers to simulate complex quantum systems such as many body interactions \cite{feynman2018simulating}. The most well-known quantum algorithms showing a quantum advantage over classical ones are Deutsch-Jozsa \cite{deutsch1992rapid,collins1998deutsch}, Bernstein-Vazirani \cite{10.1145/167088.167097}, Shor's \cite{10.1137/S0097539795293172}, Grover's \cite{10.1145/237814.237866}, Lloyd's algorithm for solving system of linear equations \cite{PhysRevLett.103.150502}, etc.  
In the present time, there is an exponential growth of data, generally referred to as big data. Data analysis or more generally data science has emerged as an area of active research aiming to unravel patterns and structures. A fundamental problem that arises with big data in various fields like pattern recognition \cite{PhysRevLett.88.018702, Diday1976}, image processing \cite{ 9605278}, machine learning \cite{ahuja2020}, etc is the clustering problem \cite{everitt2011cluster, kaufman2009finding}. Several classical as well as quantum algorithms for clustering have been proposed till date with varying complexity of resources. Classical clustering algorithms include K-means clustering \cite{Hartigan1979KMeans}, Density-based spatial clustering of applications with noise (DBSCAN) \cite{10.5555/3001460.3001507}, etc. The first quantum clustering algorithm was developed by \textit{Durr et al.} \cite{durr2006quantum}. Further, a quantum-game-based clustering algorithm was developed by \textit{Li et al.} \cite{li2009novel} along with another quantum algorithm using quantum walks \cite{li2011hybrid}. \textit{Yu et al.} \cite{yu2010quantum} proposed a quantum clustering-based algorithm for multi-variable nonlinear problem. Several other algorithms that uses quantum mechanical laws are present in the literature \cite{10.1145/1273496.1273497}, although they suffer from the requirement of a quantum black box with unknown complexity, as the algorithms usually use variations of Grover's algorithm \cite{10.1145/237814.237866}. Some recent works have also studied how quantum methods can provide alternative clustering approaches, such as, probabilistic quantum clustering approach \cite{CASANAESLAVA2020105567}, variational algorithms based quantum-inspired clustering \cite{bermejo2023variational}, quntum k-means clustering \cite{Ahsan2019, kavitha2023quantum}, quantum spectral clustering based on unsupervised learning algorithms \cite{li2022quantum}, quantum clustering algorithm for multi-dimensional datasets \cite{Luca2024} etc. 

In this work, we propose two novel measurement-based quantum clustering algorithms. Both algorithms are polynomial in time in terms of data to cluster, and the distance between the furthest points. Further, they effectively remove the requirement of a black box. The first algorithm proposed in this paper follows a divisive approach, which we call Quantum Hierarchical Clustering Algorithm (QHCA). The second algorithm is a Unsharp Measurement-based Clustering Algorithm (UMCA), where clustering of a data set is achieved by measuring the data points unsharply and with the appropriate choice of the variance of the Gaussian window. For both the algorithms, the operation complexity was found to be of the order of $O(D_{max}^2)$, where $D_{max}$ is the maximum distance between any two points in the dataset, upto some linear factor.  We show in this work, $D_{max}$ can be suitably scaled down to $O(log_N)$, with high accuracy in the clustering outcome.  Compared to other quantum clustering algorithms, we find that our proposed methods provide a similar complexity, with the benefit of easy implementability on noisy intermediate scale quantum (NISQ) devices. We further implemented the algorithms on a concentric circle data set for which the classical divisive clustering approach fails and show that our algorithm perform well in accuracy with very low qubit cost and high measurement complexity. We further show that our algorithms work with very low qubits and measurement complexity, and is thus suitable for NISQ computers. We have implemented QHCA on the Churrtiz data set of $130$ cities, and shown that the algorithm works well with decreasing qubit resources.  Additionally, UMCA is implemented on the labeled Wisconsin breast cancer dataset of $699$ datapoints, and where we found an accuracy of approximately $94\%$, using only $4$ qubits and measurements polynomial in qubit numbers. Subsequently we show, under an assumption of measurement error, UMCA works better than the noiseless measurement, which makes it extremely suitable for NISQ devices. 

The manuscript is organized as follows. In section \ref{sec 2}, we give an overview of hierarchical clustering and unsharp measurements in quantum mechanics. In section \ref{sec 3}, the workflow of both algorithms is described. Section \ref{sec 4} is dedicated to the implementation of our algorithms on various bench-marking datasets. Finally, we conclude in section \ref{sec 5} with future plans.

\section{Prerequisites}\label{sec 2}
In this section, we briefly review the standard clustering approaches and the basics of unsharp measurements.
\subsection{Classical Clustering approaches}
Clustering refers to the grouping of data points having similar characteristics. Widely used clustering algorithms include k-means, DBSCAN, mean shift clustering, and Hierarchical Clustering. How the data points are clustered differs in each algorithm. Here, we will primarily focus on hierarchical clustering, as one of our proposed quantum algorithms uses a similar approach. It is a method of cluster analysis that allows to build us a tree structure from data similarities. When hierarchical clustering follows a bottom-up approach, it is called agglomerative clustering where all the data points are treated as a single cluster which is used to form bigger clusters based on similarity of data. The process is repeated until a complete single cluster is formed. Another way to cluster the data using hierarchical clustering is through a divisive approach. It follows a top to bottom approach. Divisive clustering is done by recursively splitting a larger data set $C$ into two smaller sub-datasets until the required number of clusters is obtained. Such a method is good at the identification of large clusters and it is common in the field of data mining, image segmentation, decision-making, etc. Clusters obtained by this approach are presented as a hierarchical binary tree, which makes it attractive in many real world problems,  such as indexing problems. Divisive clustering can be used to split the data set into smaller ones until all the clusters contain only a single element, where the hierarchy in a data set of $N$ objects is built in $N-1$ steps, with $2^{N-1} - 1$ possibilities to split the data into two clusters. Whereas, in the agglomerative method, clustering of any two data points together leads to $N(N-1)/2$ possible combinations. Agglomerative clustering algorithms have the time complexity of $O(N^3)$ as we have to scan the $N \times N$ distance matrix to get the lowest distance in each $N-1$ iteration. This can be reduced to $O(N^2 \text{log}N)$ by using priority queues. By some optimizations, it can be brought down up to the order of $O(N^2)$ \cite{day1984efficient}. Divisive clustering with an exhaustive search has a time complexity of $O(2^N)$. 
 To combine (in agglomerative) or divide (in divisive) the data sets, a measure of (dis)similarity between the data points should be considered. For any two data points $x \in C$ and $y \in C$, it is given by a non-negative real-valued distance matrix $D_{xy} = d(x, y)$ where each datapoint $x \in C$ has $k$ attributes given by the tuple $x = (x_1, x_2, ..., x_k)$. The \textit{Minkowaski} metrics are the family of metrics that quantify the measure of such dissimilarity. For a fixed $p \geq 1$ and for any two data points $x$ and $y$, it is given by, 
 \begin{equation}\label{eqn:1}
     L_p(x,y) = \left( \sum_{i = 1}^k |x_i-y_i|^p\right)^{\frac{1}{p} }
 \end{equation}
 for $p = 2$, it becomes \textit{Euclidean} metric. Usually, Euclidean is used as a dissimilarity measure between two data points. 
 
\subsection{Quantum Clustering approaches}
The first quantized clustering algorithm was proposed by Durr et. al. \cite{durr2006quantum}, although it was not developed for answering the clustering problem. The algorithm calculated the quantum query complexity of graph problems and showed that the query complexity of the minimum spanning tree is of the order $O(N^{3/2})$ in the matrix model and $O(\sqrt{MN})$ for the array model in which $N$ and $M$ represent the number of vertices and edges respectively in the graph. After computing the minimal spanning tree of a graph, the data points can be grouped into $k$ clusters by removing $k-1$ longest edges of the given graph. The classical query complexity for the matrix model is known to be of the order of $O(N^2)$. Further, they showed that their clustering algorithm based on a minimal spanning tree is close to optimal, i.e. no other algorithm, classical or quantum can do clustering in better time than $O(N^{3/2})$. In the quantized versions of clustering via minimum spanning tree, divisive clustering, and $k$-medians, it turns out that they are faster than their classical analogues \cite{aimeur:hal-00736948}. For the quantum $k$-medians algorithm, the run-time was found to be $O(\frac{1}{\sqrt{{k}N^{3/2}}})$ for one iteration, which is $\sqrt{N/k}$ times faster than the classical approach. Quantum divisive clustering has a run-time of $O(N \text{log} N)$. In the quantum-version construction of the $c$-neighbourhood graph, where $c$ is the number of closest neighbors, the time complexity is $O(dN \text{log} N)$, where $d$ is the dimensionality of the space in which the data points live and N is the number of data points. However, these quantum clustering algorithms make use of a unitary black-box, which is harder to implement, whereas, the algorithms present in this work are easy to implement, with a fixed set of two-qubit gates. We provide a table comparing the time complexity and implementability of quantum clustering algorithms in Table~\ref{table}.

\begin{table}[ht]
\centering
\begin{tabular}{p{4 cm}|p{4 cm}}
\hline
\hline

\textbf{Algorithm} & \textbf{Operational Complexity} \\
\hline
Quantum Minimal Spanning tree \cite{durr2006quantum} & $O(N^{3/2})$  \\
\hline
Quantum $K$-medians \cite{aimeur:hal-00736948} & $O\left(\frac{1}{\sqrt{kN^{3/2}}}\right)$  \\
\hline
Quantum Divisive Clustering \cite{aimeur:hal-00736948} & $O(MN\log N)$ \\
\hline
Quantum $K$-Means \cite{Ahsan2019} & $O(\sqrt{M}NK)$ \\ 
\hline
QHCA & $O(KN(\log N)D_{max})$ \\
\hline
UMCA & $O(KN(\log N)\sqrt{D_{max}})$ \\
\hline
\hline
\end{tabular}
\caption{Comparison of time complexities of various clustering algorithms. $N$, $M$ and $K$ stands for number of datapoints, dimension of one datapoint, and number of required clusters. }
\label{table}
\end{table}

\subsection{Unsharp Measurements in Quantum Mechanics}
In quantum mechanics, the process of measurement is mathematically described by operators. For a given quantum state \( |\psi\rangle \), the probability of measuring a particular eigenvalue \( a \) associated with an operator \( \hat{A} \) is given by \( P(a) = |\langle \psi_{a} | \psi \rangle|^2 \), where \( |\psi_{a}\rangle \) is the eigenstate of \( \hat{A} \) with eigenvalue \( a\). This fundamental probabilistic nature of quantum mechanics highlights its stark differences from classical physics, where measurements do not generally alter the state of the system being measured. Quantum measurements are described by the collection of $\{M_i\}$ measurement operators, where the index $i$ refers to the outcome of the measurement. Assuming a measurement associated with the measurement operators $M_i$ is acted on a quantum state $|\psi\rangle$, we can define another operator $E_i^{jk} = M_i^{jl}\ M_i^{lk}$ which obeys $ \sum_i E_i = I.$ The set of operators $\{ E_i \}$ is referred to as POVM (positive operator valued measure) and the operators themselves are referred to as POVM elements.
In general, the POVM elements need not to be orthogonal to each other. The set of operators $E_i$ is enough to determine the probabilities of different outcomes of the measurement. The special case of generalized measurements, where the measurement operators are orthogonal to each other is referred as projective or sharp measurement. The eigenvectors of such operators form an orthonormal basis set for the Hilbert space and the outcome of the measurement corresponds to one of the basis states. These measurements are called projective or projection valued (PV) measurements in the sense that the initial state of the system $\ket \psi$ is projected onto one of the eigenstates $\ket{A_i}$ of an observable \textit{A} that is being measured. These measurements are called sharp as the system completely collapses to one of the eigenstates of the observable, destroying the initial state. The sharpness of measurement is intrinsically present in the mathematical structure of an observable. However,  in practice, no accurate measurement of a system is possible. For a classical system, we can predict the outcome, and hence the measurement is sharp but when dealing with real systems, quantum events are fuzzy and measurements are not accurate, also often termed as \textit{unsharp}. The unsharp measurements are represented by \textit{Effect Operators} or simply effects. One can define any quantum unsharp event as a weighted average of quantum sharp events. The event of quantum measurement can also be thought of as a self-adjoint operator having an eigenspectrum between 0 to 1. In this extended framework, we define a positive operator valued measure (POVM), which consists of certain positive operators that are self-adjoint with eigenspectrum lying in the interval [0,1].  This leads to the definition of a quantum effect as a self-adjoint operator $E$ having eigenspectrum lying between $0$ and $1$ \cite{foulis1994effect}. The effects are positive bounded self-adjoint operators such that $0 \leq E \leq I$. To give an example, For an infinite dimensional Hilbert space with continuous basis set $\{\ket x\}$ with orthogonality condition $ \bra{x}\ket{x'} = \delta_{xx'}$ an effect can be defined as, 
\begin{equation}\label{eqn:2}
    E_y = \int_{-\infty}^{\infty} dx\frac{1}{(\sqrt{2\pi \Delta^2)}} e^{\frac{-(y-x)^2}{(2\Delta^2)}}\ket x \bra x ,  
\end{equation}
where $\Delta$ is the standard deviation of the distribution. It describes an imprecise measurement of position $y$. All the effect operators obey the completeness relation. The operator $E_y$ represents the unsharp measurement of the position at point $y$. These operators indeed create a set of effect operators because they follow completeness relations and are positive operators.

\section{Quantum Clustering Algorithms}\label{sec 3}
In this section, we present our measurement-based clustering algorithms. In the first algorithm, we make use of Quantum entanglement and parallelism for clustering. In the second Quantum-inspired algorithm, the notion of unsharp measurements is used. 

\subsection{Quantum Hierarchical Clustering Algorithm (QHCA)}
The similarity measure is a key factor in constructing clustering algorithms. A measure of distance is defined over the features of data points to perceive how similar the two data points are.  
We consider a dataset consisting of $N$ elements with each element having $d$ attributes. The measure of dissimilarity between any two data points is given by Euclidean distance metric as in Eq.~\ref{eqn:1} with $p = 2$. Data points having similar distance measures are clustered together.  We now present our algorithm for clustering the data points based on the distances between them.

\begin{center}
\begin{figure}[!htb]
     \centering
     \includegraphics[scale=0.2]{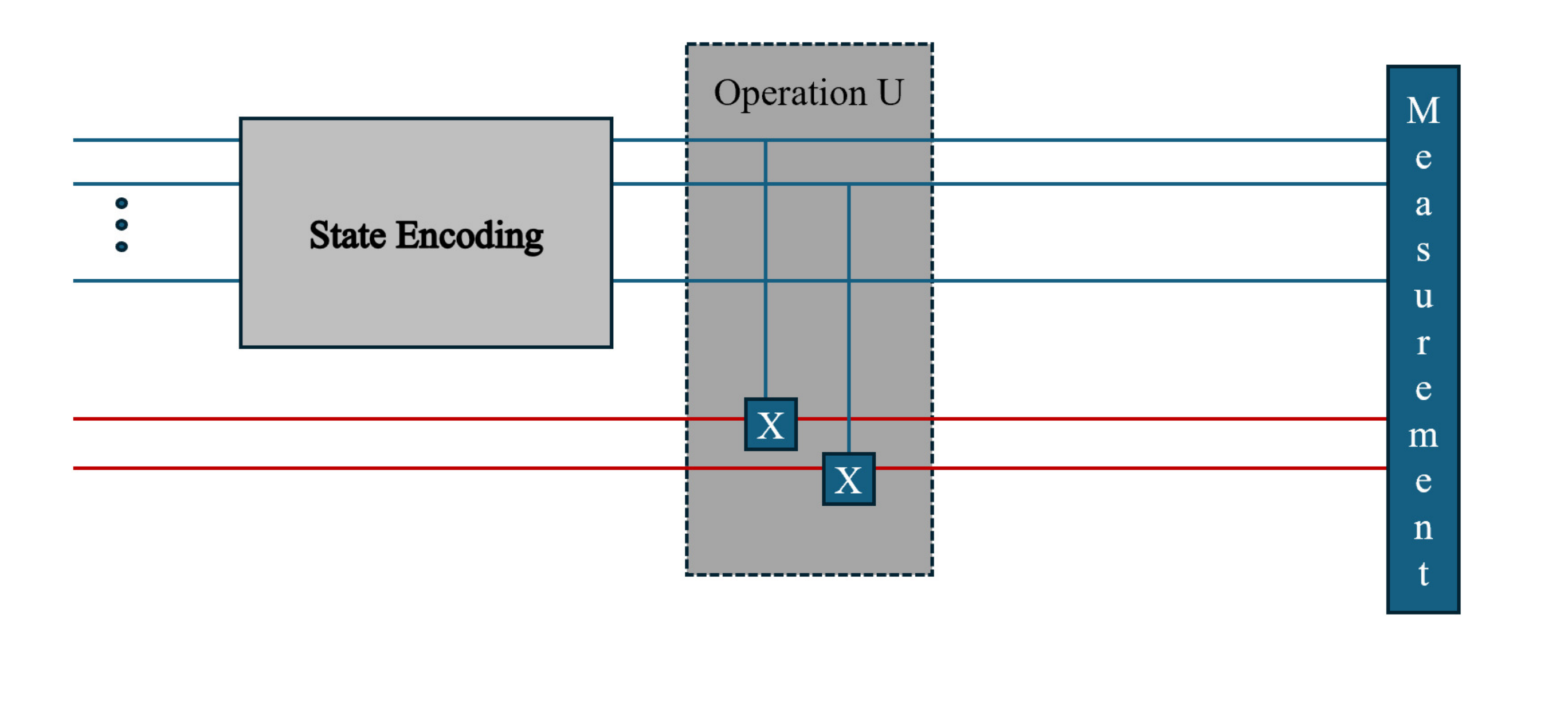}
     \caption{Circuit representation of QHCA. The blue register represent the quantum register that encodes the distances. The red register is an ancilla register, with number of qubit = $log_{2} \text{(Number of clusters)}$.}
\label{Fig:Data2}
\end{figure}
\end{center}

\begin{figure}
\centering
\includegraphics[scale=0.3]{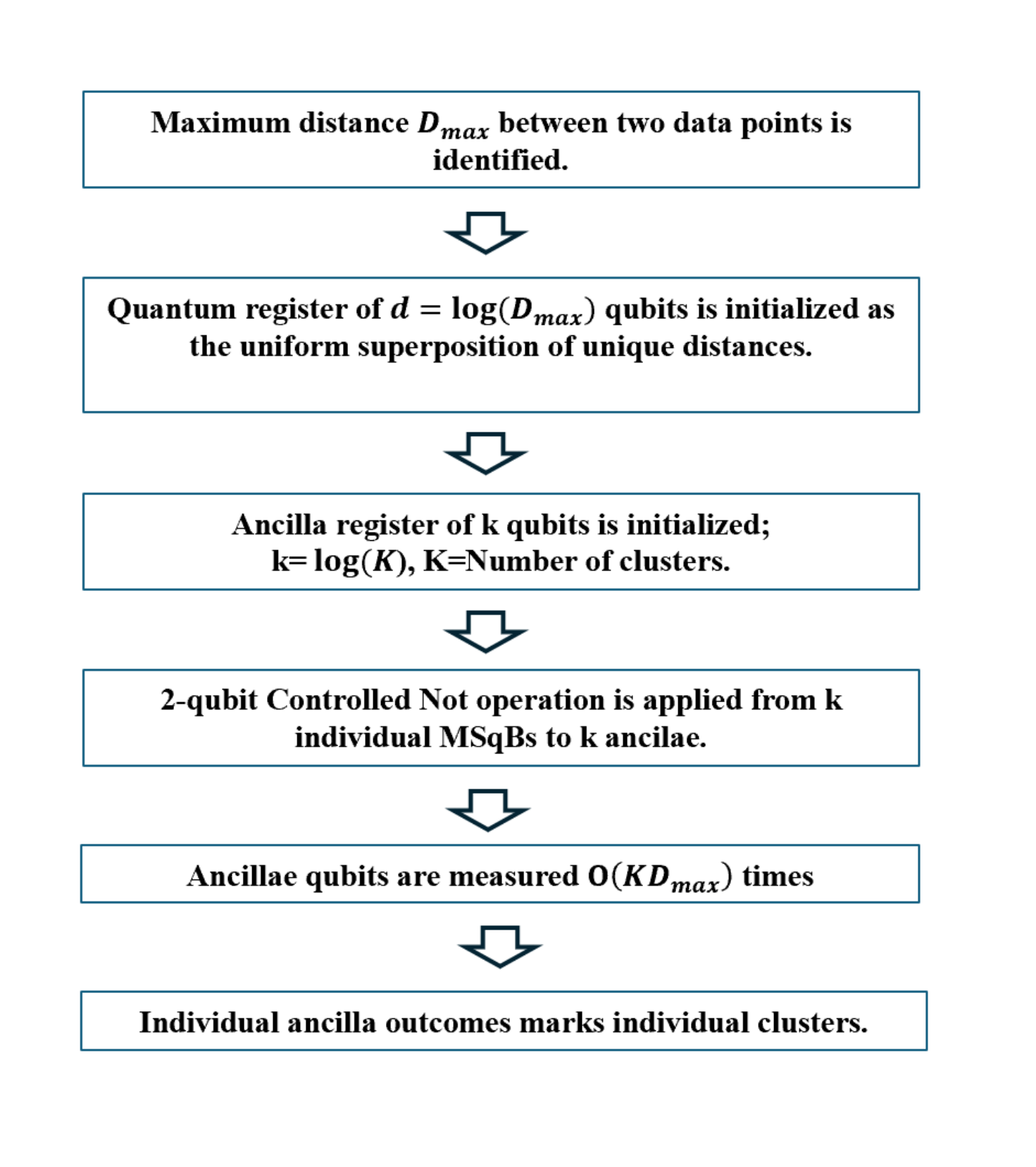}
     \caption{Flowchart of QHCA to prepare $K$ clusters.}\label{Fig:Data3}
\end{figure}

\begin{itemize}
\item We first run a classical sorting algorithm on the entries of the dataset $C$ and choose the two data points $k_1$ and $k_2$ that are farthest from each other, i.e., the distance between $k_1$ and $k_2$, $D_{max}$ is the maximum distance between any two pairs of points. This process generally requires $O(N^2)$ comparisons in the classical domain, although, using the subroutine proposed in \cite{aimeur:hal-00736948} we can substantially reduce this cost to the order of $O(N)$.
  
\item We choose $k_1$ as the origin and redefine the distances of the $N$ entries from $k_1$ in terms of the binary equivalent of their closest integer. This gives a way to represent the distances as $d-$qubit quantum basis vector states, where $d=\lceil \log_2D_{max} \rceil$. As an example, if the distance between point $k_1$ and any other point $k_i$, where $1 \leq i \leq N$ is given by $C_{k_1i}$, we represent it as $\ket{D_{k_{1i}}}$, where $D_{k_{1i}}$ is the binary equivalent of the decimal number $\lfloor C_{k_1i}\rceil$, i.e., the closest integer to $C_{k_1i}$.  The best time complexity known is $O(\log N)$ with the space complexity of $O(\log N)$. 

\item We now take a $d$ qubit quantum register, where $d= O(\lceil \log_2D_{max} \rceil)$, and represent the distances on this register as an equal superposition state $\ket{\psi}$ of all quantum states representing the individual distances $\ket{D_{k_{1i}}}$. We emphasize that the state $\ket{\psi}$ have each distance value only once, i.e., data-points with similar distances are represented by only one basis state. Mathematically $\ket{\psi}$ is written as, 
    \begin{equation}\label{psi}
        \ket{\psi} = \frac{1}{\sqrt{N^{\prime}}}\sum_{i = 1}^{N^{\prime}}\ket{D_{k_{1i}}}, 
    \end{equation}
where $N^{\prime}$ is the number of basis vectors present in the register, and  $N^{\prime} \leq N$. Further, the number of qubits $d$ is in the order of $log_2 D_{max}$. We further allow a scalar multiplication of the distances between the data points, if necessary, to maintain a balance between a low value of $d$, and accuracy of clustering. 

There are several ways of preparing the quantum state $\ket{\psi}$. The most general way of preparing this state is to implement a Grover oracle, with operation complexity $O(\sqrt{\frac{\lceil D_{max} \rceil}{N^{\prime}}})$ \cite{10.1145/237814.237866}. Depending on the number as well as nature of basis states $N^{\prime}$, this state can be prepared with significantly less operations \cite{Shukla2024, 9506863}.

The quantum state representing the superposition of distances is designed in a way that when the most significant qubit (MSqB) in its register, qubit $d$, is in state $\ket{1}$; it represents the distances of data points farthest from the origin $k_1$. Approximately these distances are $\geq \frac{D_{max}}{2}$. On the contrary, when qubit $n$ is in state $\ket{0}$, it represents the distances of from $k_1$ that are approximately $\leq \frac{D_{max}}{2}$. Similarly, it can be easily seen that when qubit $n-j$ is at state $\ket{0}$ and at $\ket{1}$, it approximately represents the distances  $\leq \frac{D_{max}}{2^{j}}$ and $\geq \frac{D_{max}}{2^{j}}$ respectively.  
\item We need to prepare the clusters such that they are within a pre-fixed distance $D_{min}$ from their individual centers; alternatively, we can also create a predefined number of clusters, say, $K$. We take $m=\lceil \log_2\frac{D_{max}}{D_{min}}\rceil$ ancilla qubits in the first case, and $m=\lceil \log_2 K\rceil$ ancilla qubits in the second case, and apply the operation $U$ on the register $n+m$, given as, 
\begin{align*}
    U = \Pi_{i_n=0, i_m=0}^{m,m} C^{n-i_n}X_{i_m},
\end{align*}
where $i_n$ is the $n-i_n$th MSqB in the distance register, and $i_m$ is the $i_m$th MSqB in the ancilla register. Simplistically, this operation is equivalent to applying two-qubit controlled-Not gates from the MSqBs to the ancillas. The operation complexity of this part is logarithmic in $K$. 

\item As shown in Figure~\ref{Fig:Data2}, upon measuring both the registers together, one can easily select the states of the ancilla register and find the clusters associated with it. This procedure is repeated until we get the desired number of clusters. To characterize all possible clusters through measurement, one needs to repeat the whole procedure $O(\lceil \frac{D_{max}^2}{D_{min}}\rceil ))$  ($O(KD_{max})$)times, as there might be some clusters with exponentially less data points in it. This means, it will take exponential measurements in qubit numbers, i.e., 
$O(2^{d+m}$), and linear measurement in terms of the maximal distance and number of clusters. The overall time complexity of the QHCA is thus $O(\lceil \frac{D_{max}^2}{D_{min}}\rceil N\text{log}N)$ ($O(KD_{max}N\log N)$), whereas the operational complexity is $O(k\sqrt{\frac{\lceil D_{max} \rceil}{N^{\prime}}})$ combining the state preparation and clustering part.  
\end{itemize}

\subsubsection{Deciding the boundaries}
The boundary of each cluster is determined by the number of qubits required for representing the distances and the number of qubits required for clustering. If we consider $k$ clusters to be formed and the distance statevector requires $d$ qubits for the representation, then the upper and lower limit of each cluster $t$ can be explicitly found. Say, the $t^{th}$ cluster is represented by its binary equivalent number $\{t_{k-1}...t_1t_0\}$ and the bit-string that encodes the distance between the center and the cluster $d^t$, is represented as $\{d_{n-1}...d_1d_0\}$ where 
$t_i, d_k \in \{0,1\}$ and $k-1 \leq i \leq 0$, $n-1 \leq k \leq 0$.
The upper limit for cluster $t$ is given by 
\begin{equation}\label{eqn:7}
    d_{max}^t = \sum_{j=0}^{k-1} 2^{j+k}\cdot t_j + \sum_{j=0}^{k-1} 2^j
\end{equation}
 and the lower limit is given by 
 \begin{equation}\label{eqn:8}
     d_{min}^t = \sum_{j=0}^{k-1} 2^{j+k}\cdot t_j.
 \end{equation}
The schematic of the algorithm is shown in Figure~\ref{Fig:Data3}.

\subsection{Unsharp measurement-based clustering algorithm (UMCA)}
Given a set of data, we find the largest distance among any two points, and consider any one of those two points, say $k_1$. Subsequently, we represent each unique distance $k_1$ as a quantum state encoded by the binary equivalent of the corresponding closest integer. Then,  we create an equal superposition of all the states. As described in section \ref{sec 2}, A set of effect operators can be constructed corresponding to each state. The construction of each effect operator is such that, if one unsharply measures the prepared superposed quantum state using one of them, say $E_i$, where $\ket{i}$ is the quantum state at the origin of the effect  (reference: Eq.\ref{eqn:2}), the states that are nearer to $\ket{i}$ are automatically clustered together after being measured. Depending upon the width of the gaussian, one can control the probabilities of the other states near to state $\ket{i}$ in the measurement outcome \cite{Barui2024}, such that the unsharp measurement 'selects' those states with high probability that are closer to the selected origin. To divide a dataset in $K$ clusters, $K$ origins are to be selected and $K$ unsharp measurements with a fixed width should be performed on the dataset. In case of repetition of a datapoint in more than one cluster, it can be assigned to any one of them. 

In computational basis, the number of qubits required to represent all the distances is of the order $log_2(D_{max})$, where $D_{max}$ is the maximum distance between the data points. We represent individual distances as the computational basis states, and take an equal superposition of all the distances, as given in Eq.~\ref{psi}, 
\begin{equation*}
        \ket{\psi} = \frac{1}{\sqrt{N^{\prime}}}\sum_{i = 1}^{N^{\prime}}\ket{D_{k_{1i}}}, 
    \end{equation*}
where $N^{\prime}$ is the number of \textit{unique distances} present in the register, and  $N^{\prime} \leq N$. After this state is prepared, one can perform unsharp measurements for a pre-defined number of times through an Effect operator, centering around the largest distance, and a pre-defined width.

\begin{figure}
\centering
\includegraphics[scale=0.3]{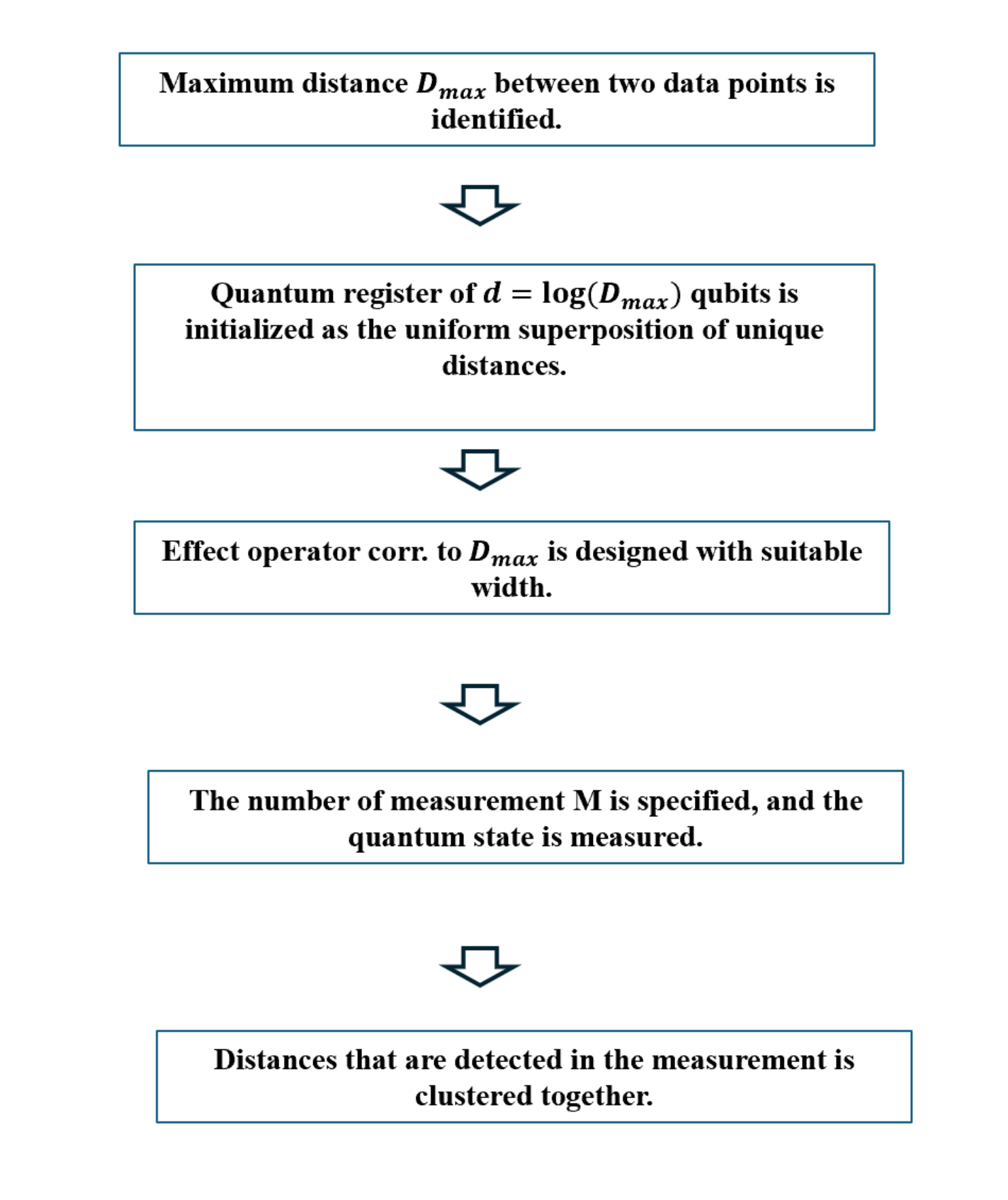}
 \caption{Flowchart of unsharp measurement based clustering algorithm.}
\label{UM_flow}
\end{figure}

For this particular case, the set of effect operators $\{E_i\}$ is defined as follows:
\begin{equation}\label{eqn11}
    E_i = \sum_{j \in \{0...00\}, j \neq i}^{\{1...11\}} \frac{1}{\sqrt{(2 \pi \Delta^2)}} e^{-\frac{(i-j)^2}{2\Delta^2}} \ket{j}\bra{j},
\end{equation}
where $E_i$ refers to the Effect operator corresponding to $i-$th distance in the quantum state $\ket{\psi}$, and $j$ is the $j-$th distance. The operator mimics a Gaussian distribution, with $\Delta$ as the standard deviation of the Gaussian, and $i$ as its center. Each Effect operator can thus be represented as a diagonal matrix with entries given by $\frac{1}{\sqrt{(2 \pi \Delta^2)}} e^{-\frac{(i-j)^2}{2\Delta^2}}$. It can also be thought of as a weighted average of POVMs $\ket{j}\bra{j}$ with weights $\frac{1}{\sqrt{(2 \pi \Delta^2)}} e^{-\frac{(i-j)^2}{2\Delta^2}}$. We keep the effect operator centered around the state representing the largest distance, as this will allow us to cluster the distances that are furthest away from the origin.

Below we present a workflow of the unsharp-measurement based clustering algorithm. 
\begin{itemize}
    \item Similar to the previous algorithm, we run a classical sorting algorithm for finding the points which are farthest apart, and choose (any) one of them as the origin. From this point, the distances of all other points are computed with the Euclidean distance. An equal superposition of state of all unique distances is then created on a quantum register with $d = log(D_{max})$ qubits.
    \item The effect operators described by Eq.~\ref{eqn11} correspond to the unsharp measurement around one particular distance, i.e., the center of the effect operator ($i$) can be selected to be one of the distances. Thus, there can be $2^d$ effect operators, each centered around a single distance. However, as we show in this work, only one effect operator is needed for effective clustering. We choose the effect operator that is centered around the largest distance, and choose the width of the Gaussian to be $\Delta=\frac{D_{max}}{k}$, where $k$ is the required number of cluster.
\item Next we measure unsharply around the specified center for a specified number of times. By design, the quantum states closer to origin will appear in the outcome with higher probability. The number of measurement will depend on both the width of the Gaussian, and the number clusters $k$.
\end{itemize}

\subsubsection*{Complexity:} The complexity of preparing the quantum register is similar to the previous algorithm, i.e., $O(N\log N)$ for classical preprocessing, and $O\sqrt{\frac{D_{max}}{N^{\prime}}}$  for state preparation. We show in Section~\ref{sec 4}, that even with polynomial (in qubit number $d$) measurements, it is possible to achieve a high accuracy with this method. To perform this measurement on a quantum hardware, aside from setting measurement aparatus in a specified manner,  one can also projectively measure the superposition of all unique distances in computational basis state, and then assign the suitable weights to them. The clusters are made according to the corresponding amplitudes of the state after the (unsharp) measurement.

\section{Algorithm Implementation }\label{sec 4}
In this section, we will demonstrate the implementation of both algorithms on three standard datasets. First, we implement the quantum clustering algorithm on the concentric circle dataset as well as the Churriz dataset of the cities. We have used both statevector simulator and the QASM simulator for the implementation of QHCA on the concentric circle dataset.  
\subsection{Implementation of QHCA on the concentric circle dataset}
\begin{figure}
\centering
  \includegraphics[scale=0.2]{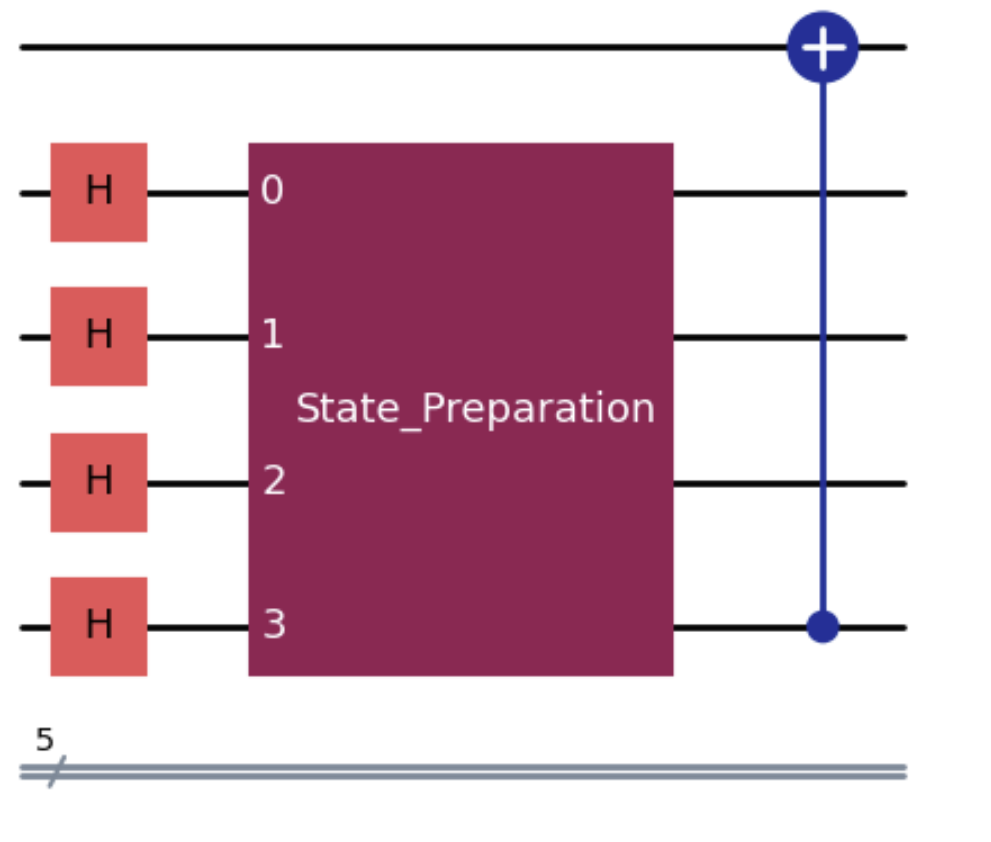}
  \caption{Circuit for clustering  the concentric circle dataset.}
  \label{conc_qhca_a}
\end{figure}

\begin{figure}
  \centering
  \includegraphics[scale=0.2]{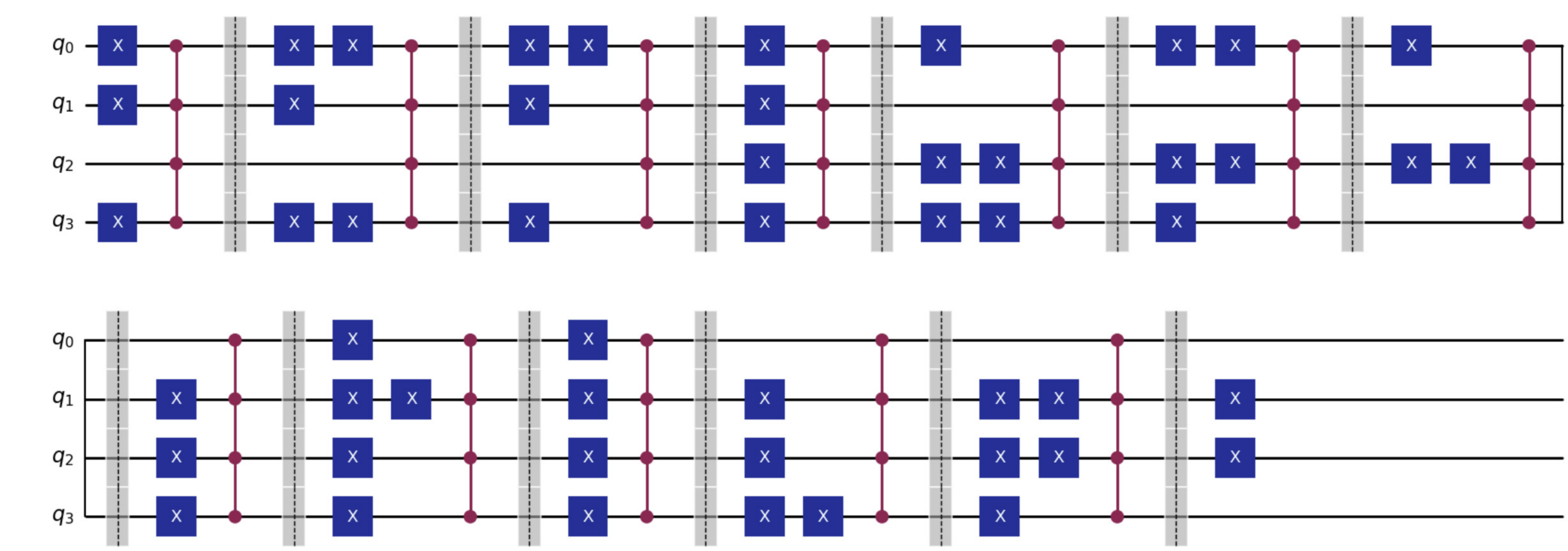}
  \caption{The 'State Preparation' block in Fig.~\ref{conc_qhca_a}. }
  \label{conc_qhca_b}
\end{figure}
\begin{figure}[H]
\centering
        \begin{tabular}{cc}
    \includegraphics[ scale=0.3]{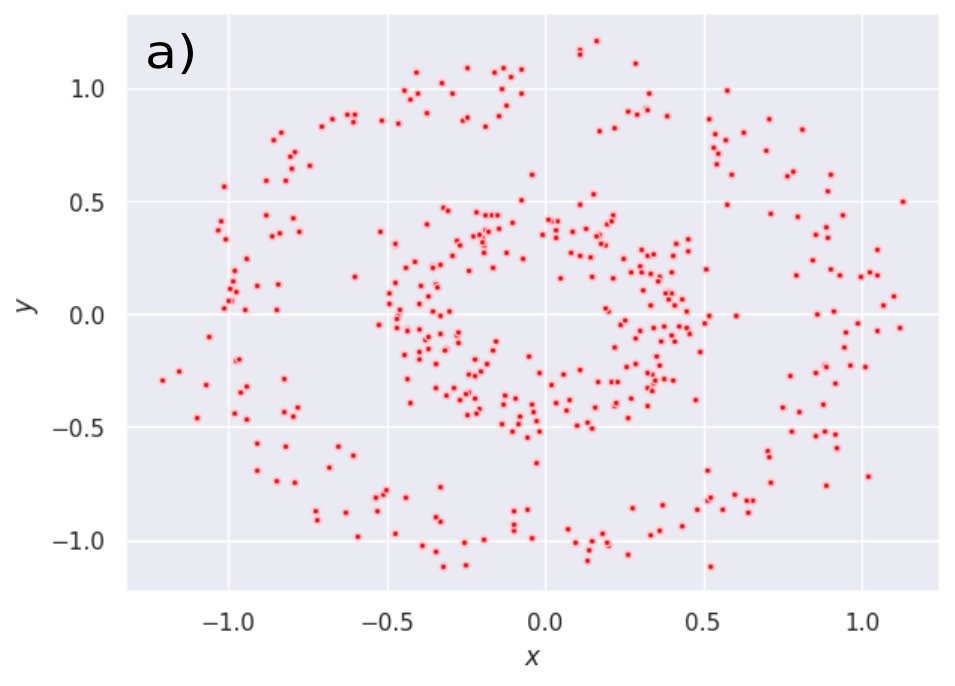} & \includegraphics[trim={1.5cm 0 0 0},clip, scale=0.3]{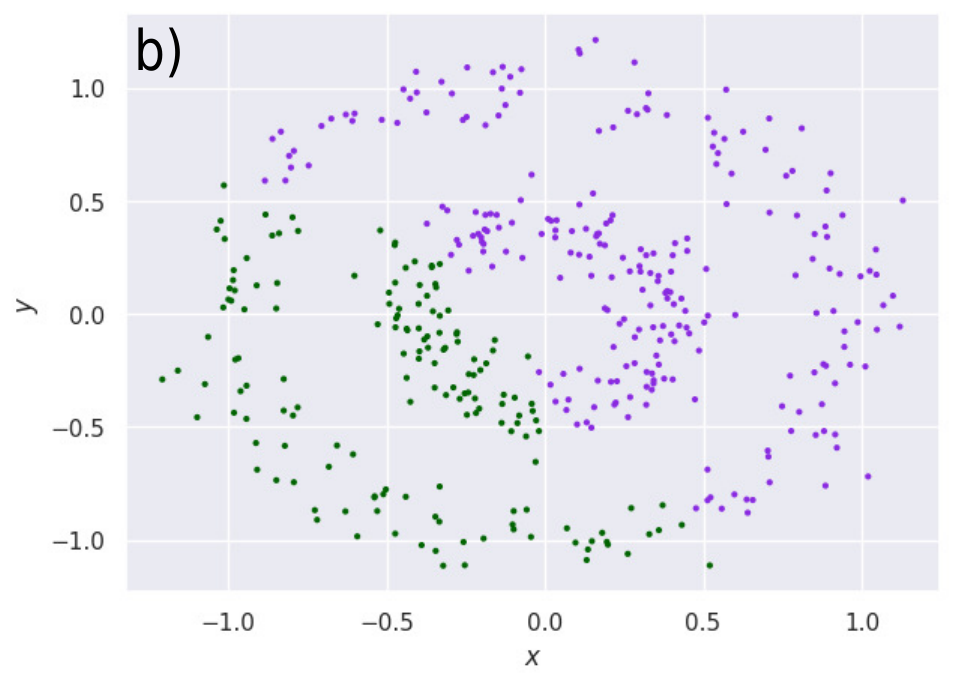} \\ \includegraphics[ scale=0.3]{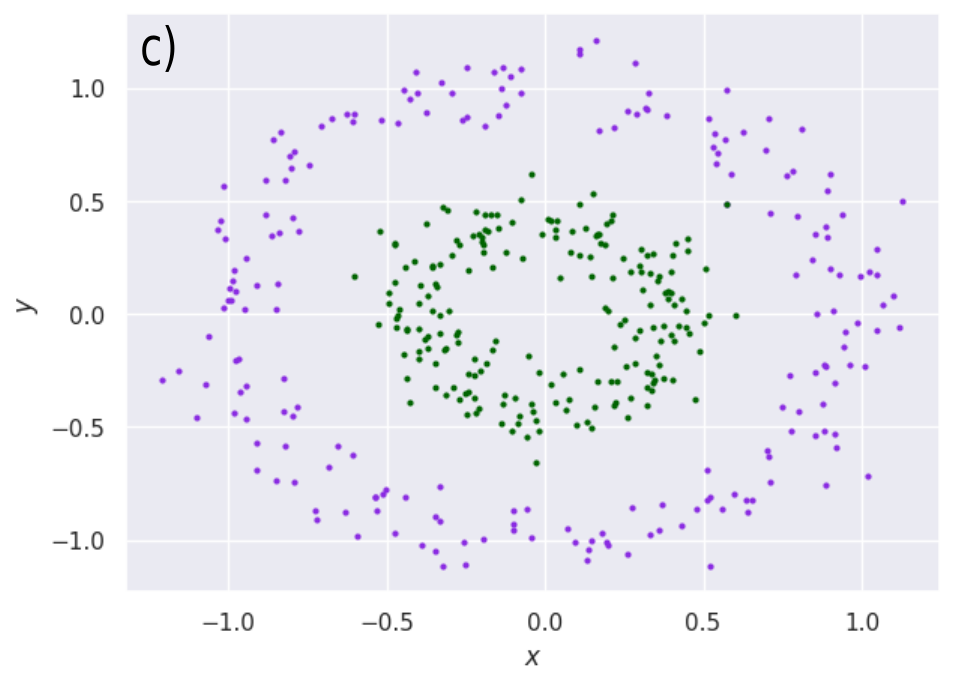}& 
   \hspace{0.5 cm} \includegraphics[trim={0 0 0 0},clip, scale=0.22]{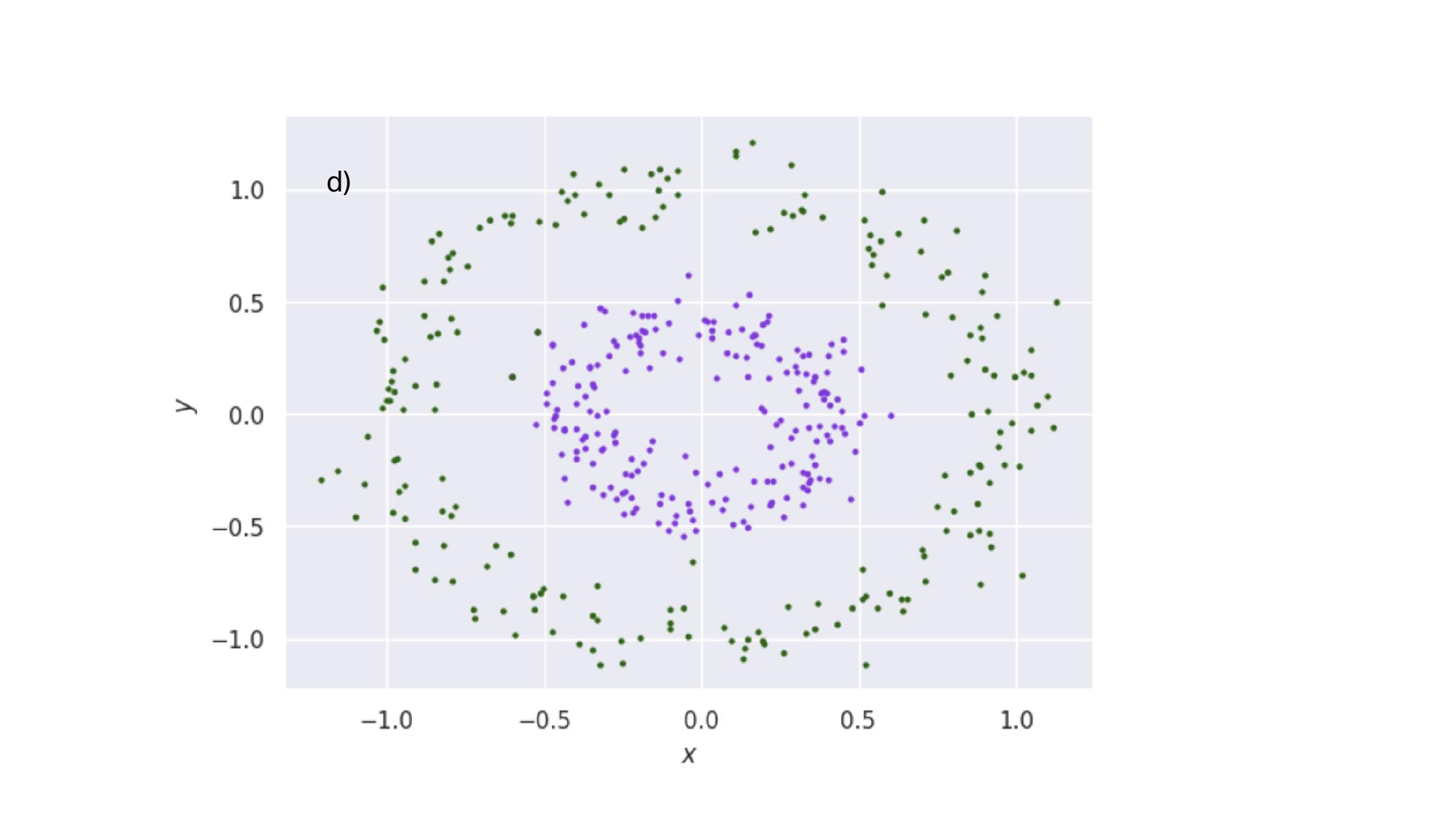}
    \end{tabular}
         \caption{(a) Dataset of 400 points generated using \texttt{sklearn} with noise ratio of 0.1 with $D_{max} = 12.48$ units. b) Classification of the dataset using the traditional classical divisive clustering algorithm. c) Classification of the dataset using the QHCA algorithm with 10000 measurements. d) Classification of the dataset using the QHCA, and using 32 measurements.  Different colors represent different clusters.}
         \label{Fig:Data6}
    \end{figure}

 We generated a dataset of 400 samples lying on two concentric circles with origin at $(0,0)$ using \texttt{sklearn} package as shown in Fig.~\ref{Fig:Data6}, and implemented our algorithm on this dataset. The data lies in the two-dimensional Euclidean plane. We first compute the maximum distance $D_{max}$ from the origin to the farthest data point, and have scaled it up with a multiplicative factor of $10$. The number of qubits required to encode all the unique distances in an equal superposition state is $n = log_2(D_{max})$. We further found despite having $400$ points in the dataset, it has only $12$ individual distances to be encoded, post the pre-processing. Once the distances are encoded in the state, we choose the number of qubits in the ancillary register to be $1$, as we want $2$ clusters here. The quantum circuit for QHCA on this particular dataset is provided in Figs~\ref{conc_qhca_a} and \ref{conc_qhca_b}. Here we have used a Grover oracle to prepared the uniform superposition of all distances, and since we have only $12$ distances in our initial quantum state, the optimal operation for Grover oracle is $O(\sqrt{\frac{12}{16}}), i.e.,  \approx 1$. We have implemented this oracle only once for clustering. Fig.~\ref{conc_qhca_a} represents the circuit to cluster the Concentric circle dataset presented in Fig.~\ref{Fig:Data6} through QHCA. Fig.~\ref{conc_qhca_b} provides the Grover oracle required to prepare the uniform superposition of distances prior to Clustering. After preparing the resister in the equal superposition of unique distances, we have applied a controlled-Not gate from the MSqB to the ancila, and measured the entire resistor to finally achieve the clusters. The whole process needs to be repeated at least $2^{4+1}$ time, i.e., $32$ times. To show the advantage of the quantum clustering algorithm over the classical counterpart, we have clustered the same data using a classical hierarchical clustering algorithm. The results are shown in Figure~\ref{Fig:Data6}. The classical algorithm fails to cluster the data correctly while the quantum algorithm does an accurate clustering. Further, we  have provided our results where the number of measurements of our circuit are 10000, and 32, and as we can see in both the cases, the quantum algorithm performs better than the classical one, however, measuring the circuit 10000 times yields slightly better result than measuring it 32 times.

Divisive clustering with an exhaustive search has a time complexity of $O(2^N)$ and agglomerative clustering algorithms have a time complexity of $O(N^3)$ as we have to scan the $N \times N$ distance matrix to get the lowest distance in each $N - 1$ iteration. The proposed quantum clustering algorithm has a time complexity of the order of $O(\lceil \frac{D_{max}^2}{D_{min}}\rceil N\text{log}N)$, which with appropriate $D_{max}$ value, provides an advantage over its classical counterpart.

 \subsection{Implementation of QHCA on Churriz dataset}

\begin{center}
    \begin{figure}[!htb]
    \begin{tabular}{cc}

    \includegraphics[scale=0.35]{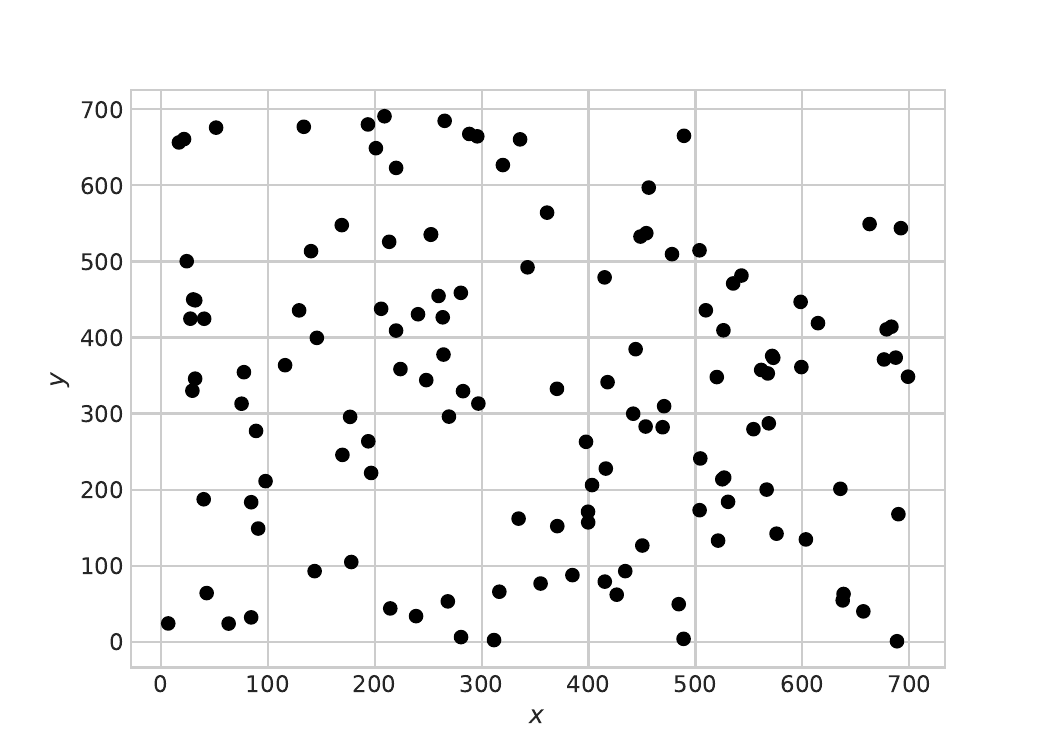} & \includegraphics[scale=0.35]{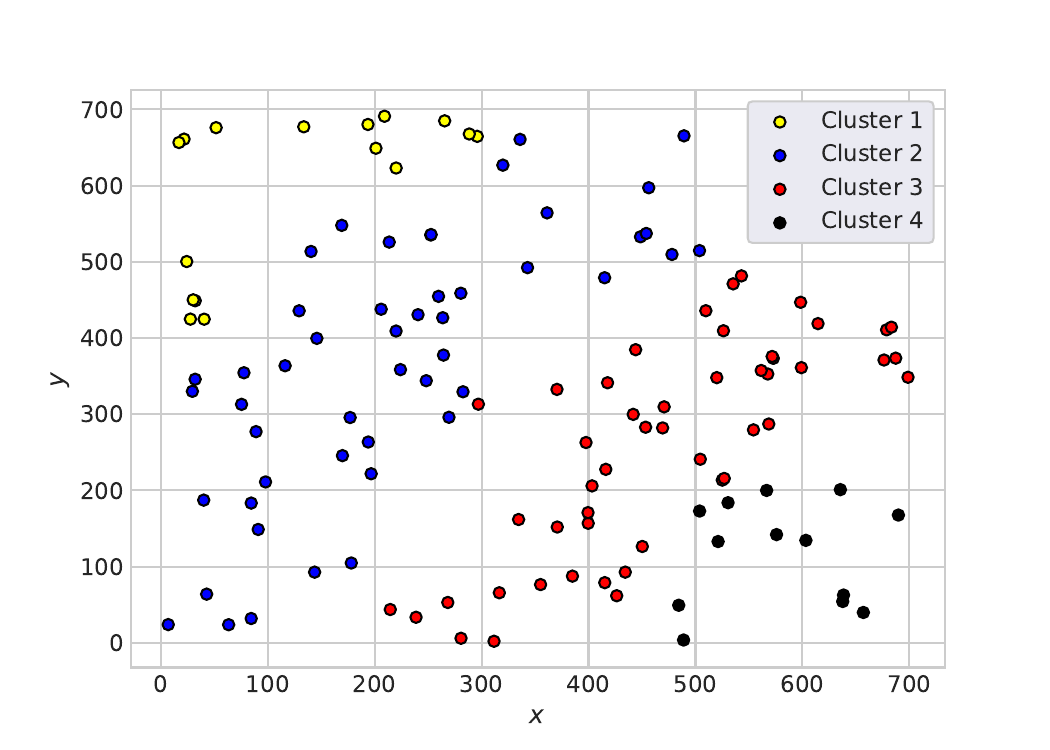}

\end{tabular}
         \caption{Churritz dataset consists of 130 cities with $D_{max} = 938.842$ kms. Each point represents the position of the city in the Euclidean plane (left) and  Clusters of nearby cities after the implementation of QHCA (right). 
         }
         \label{Fig:Data4}
    \end{figure}
\end{center}

\begin{figure}[!htb]
\centering 
    \includegraphics[scale=0.7]{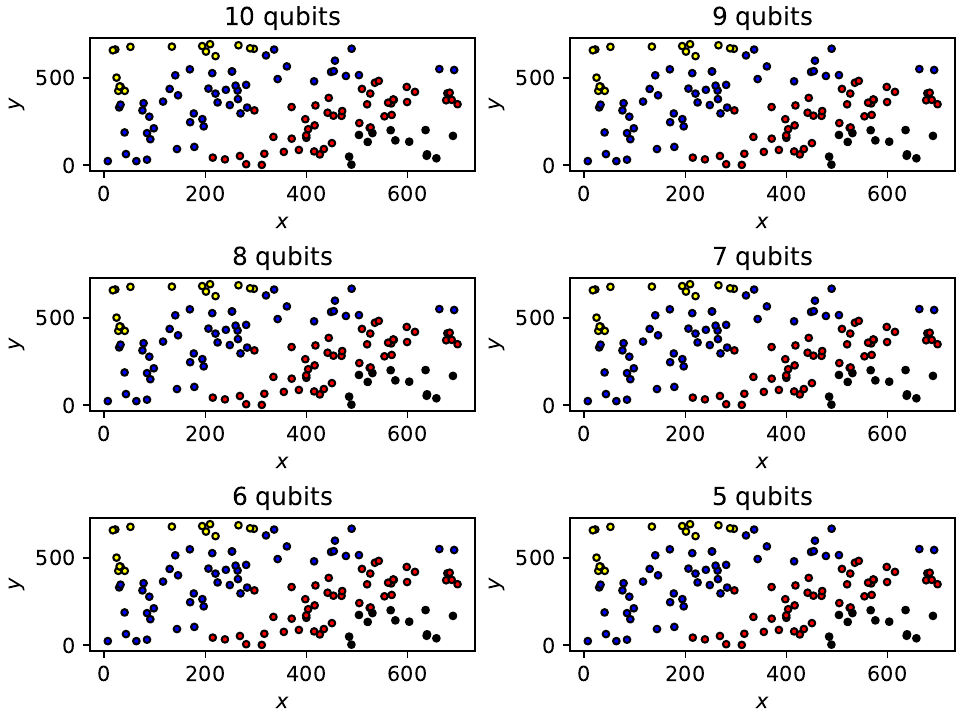} 
    \caption{Clustering of Churritz data set after scaling down the distances between cities. The clustering remains same, yet required quantum resource decreses.}
         \label{Scaling}
    \end{figure}

We use the Churritz data set of 130 cities and implement our algorithm to create $4$ clusters based on the distances. The data can be found in 
 \cite{reinhelt2014tsplib}. First, the maximum distance $D_{max}$ between any two cities is calculated, and found to be $938.842$ km. Further, we take one of the two cities that are furthest apart as the origin and cluster all the other cities accordingly. 
The dataset in the Euclidean $x-y$ plane is shown below in Figure~\ref{Fig:Data4} (left). The clustering of nearby cities is shown in Figure~\ref{Fig:Data4} (right). The number of qubits required to cluster this dataset is $10$ (to encode the distances) $+ 2$ (for clustering). 

We further scale down the distances, and check performance of QHCA on this dataset. We have scaled down the distances by diving it with increasing power of $2$  recursively, and clustered the dataset with scaled down distances using QHCA. As can be seen from Figure~\ref{Scaling}, it does not impact the clustering of the data. The clustering considering maximum distance $938.842$ km, using $10$ qubit is similar to clustering considering $29.33$ km, and using $5$ qubits. 

\subsection{Implementation of unsharp measurement-based algorithm on Wisconsin cancer dataset}

\begin{figure}[!htb]
        \begin{tabular}{cc}
    \includegraphics[scale=0.3]{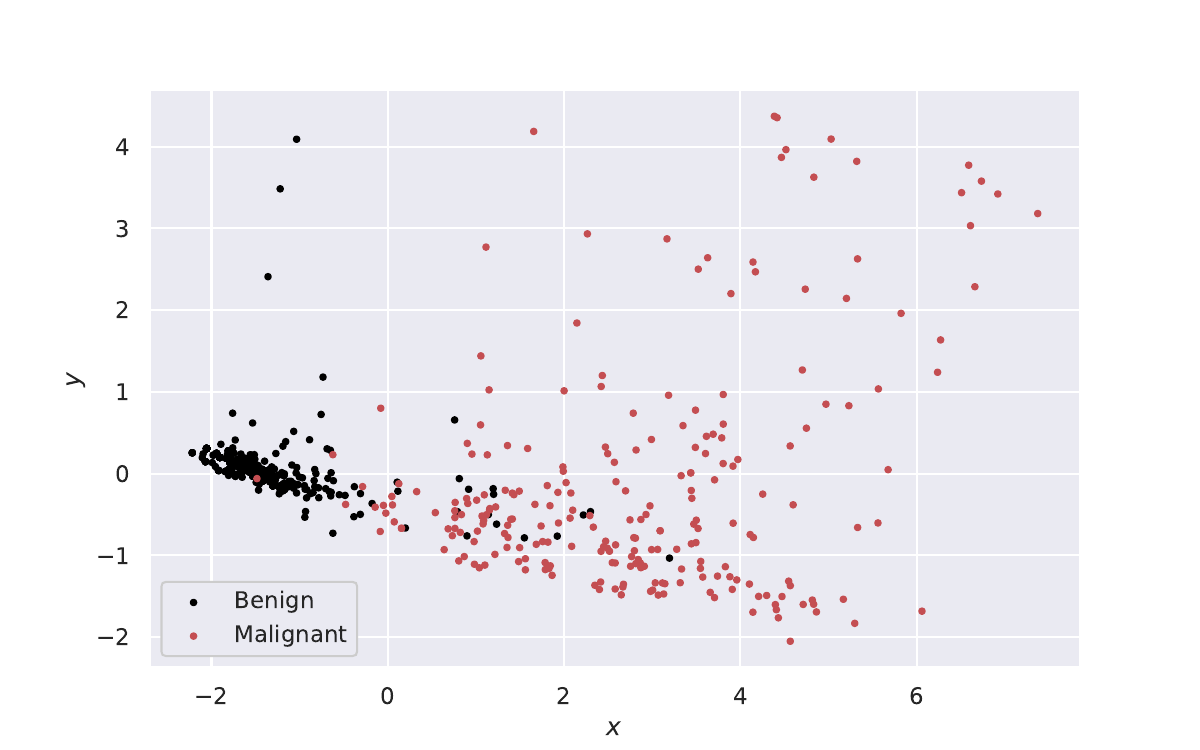} & \includegraphics[scale=0.3]{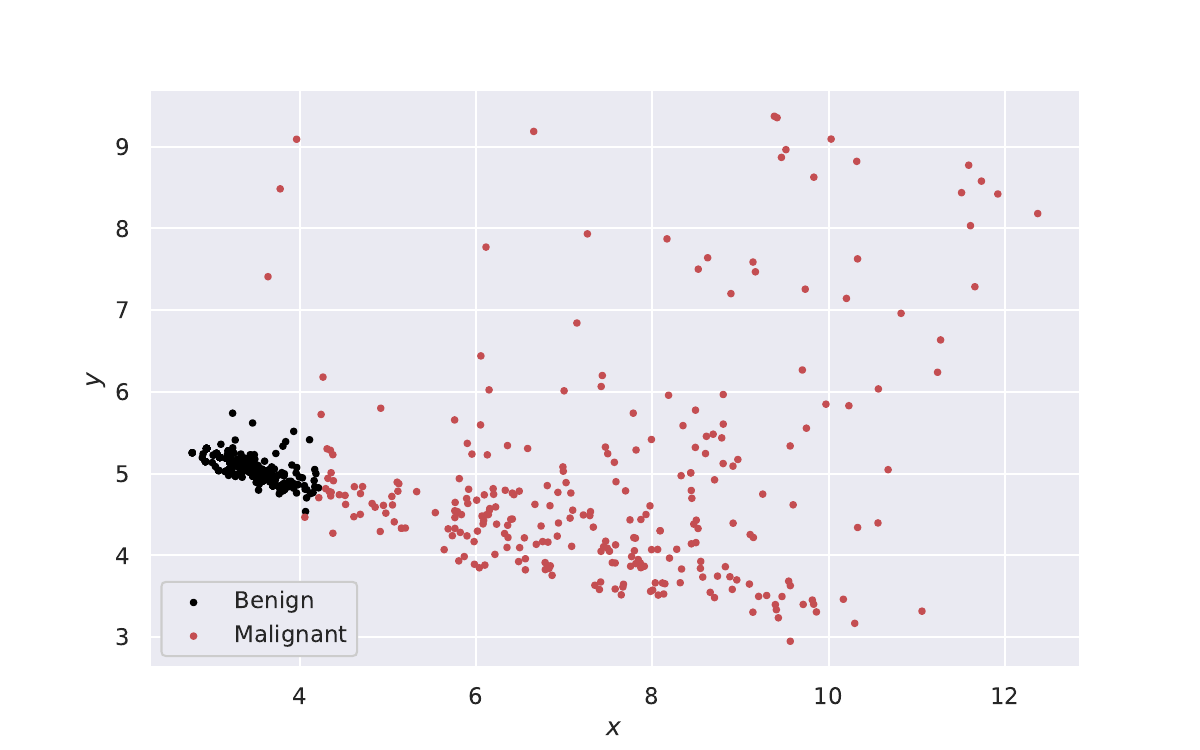}    
    \end{tabular}
         \caption{Original classification of Wisconsin Breast Cancer Dataset after feature reduction (left) and its classification with unsharp measurement based clustering algorithm (right) with $4$ qubits and $O(4^3)$ measurements. The accuracy is found to be $93.99 \%$.}
         \label{Fig:Data8}
    \end{figure}
The Wisconsin Breast Cancer dataset is a labelled dataset which consists of 699 entries with 9 different attributes such as clump thickness, cell size, cell shape, etc. In order to cluster the dataset into malignant or benign, and for better visualization, we do a principle component analysis (PCA) to reduce the features to 2 dimensions. The left hand side picture in  Figure~\ref{Fig:Data8} shows the original classification of the data. Further,  we use our unsharp measurement based clustering algorithm to classify the dataset. 
The largest distance between the points is found to be $10$, and only $4$ qubits are needed to classify the dataset. The uniform superposition of unique distances is created. For this particular case, this state is found to be,
\begin{equation}
    \ket{\psi_{wbc}} = \sum_{i=\ket{0000}}^{\ket{1010}} {\ket{i}},
\end{equation}  
implying, for this particular case, one can design this state using $O(log_{2}10)$ gate complexity and circuit depth \cite{Shukla2024}.
\begin{figure}
\centering 
\includegraphics[scale=0.60]{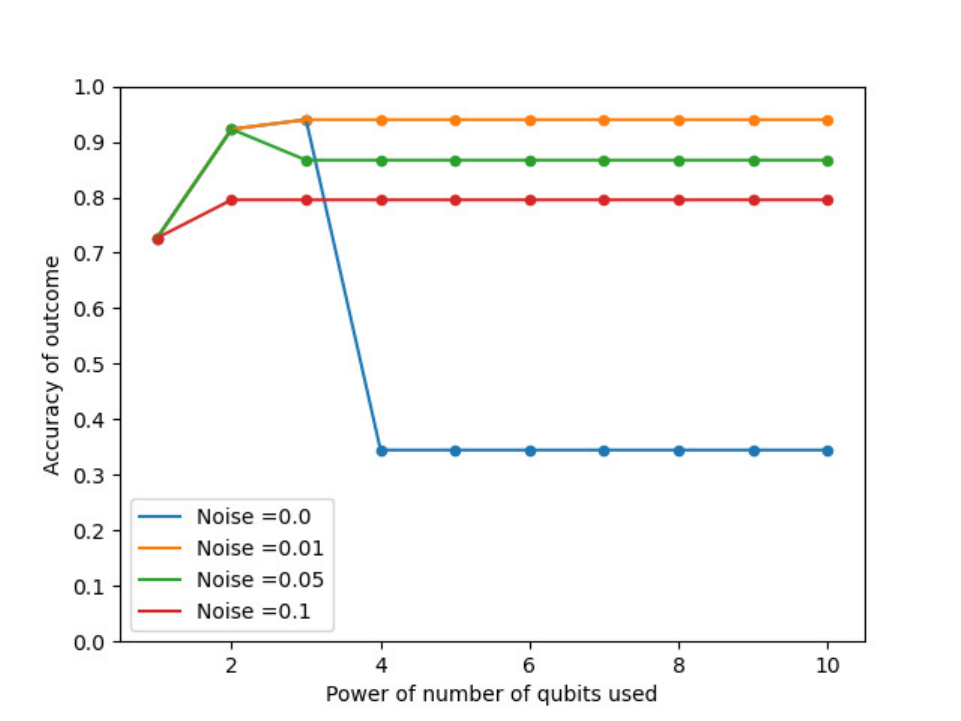}
\caption{Accuracy of the quantum unsharp measurment based clustering protocol for different measurement settings on the Wisconsin breast cancer dataset.  The X-axis represents the polynomial power of the number of qubits in the classification limit, where the Y-axis represents the accuracy of the protocol. The blue curve assumes a noise free measurement. Different colors of the curve represent different measurement settings, based on assumed noise level in measurement.}
\label{UM_noise}
\end{figure}
 
We choose the center of the Gaussian effect operator to be basis state $1010$, as it represent the maximum distance. SInce we will be performing a binary classification, we keep the width of the gaussian to be $\frac{10}{2}=5$. Finally, we choose the number of measurements $M$ one need to apply on the state  $\ket{\psi_{wbc}}$ for effective clustering. According to the protocol, the states that are detected in the measurement, are clustered together. For this protocol, we have kept $M$ strictly polynomial in $d=4$. We have varied $M$ from $d^1$ to $d^{10}$, and found that for $M=d^3$, the protocol can classify the dataset into benign and malignant classes with $93.99\%$ accuracy. The classification of the dataset with these parameters is shown in the right hand side image of Figure~$\ref{Fig:Data8}$. 

We further present the varying accuracy of the protocol under different measurements and clustering settings in Figure.~\ref{UM_noise}.  The blue curve in Figure.~\ref{UM_noise} represents an ideal scenario, where all the states that are detected in measurement, are clustered together. As can be seen, the accuracy of the protocol increases initially as one increases the number of measurements, and accuracy attains its maximum when the state $\ket{\psi_{wbc}}$ is measured $4^3$ times, however, as the number of measurements increases further, the accuracy decreased, and saturated at $34\%$. This is anticipated by design, as the effect operator is devised to measure all state with some (however small) probability.  We propose a method to combat this situation; where instead of assigning all states that are present in the measurement outcome, we only choose the states that are present with probabilities above a certain limit. For this problem, we choose this limit to be $m_{\epsilon} = \epsilon \times d^k$, where $\epsilon$ is a small number between $0$ and $1$. We present the results for three different values of $\epsilon$ in Figure~\ref{UM_noise}, depicted as 'noise', as the quantity $m_\epsilon$ is analogous to the error in measurement due to the noise present in the measurement setting of a quantum hardware. As can be seen from  Figure~\ref{UM_noise}, the  algorithm with a limit on clustering probability works best when $\epsilon=0.01$; which corresponds to measuring the state with any polynomial power of the qubit number and in general better than the 'ideal' case. We report that for $\epsilon \in [0.01, 0.1]$, the accuracy remains within the red and yellow curves as shown in Figure~\ref{UM_noise}, and for $\epsilon \geq 0.1$, the accuracy decreases, as in this case, the limit of clustering $m_{\epsilon}$ increase too much, and lesser number of basis states are clustered together.   

Even when one considers $m_{\epsilon}=0.0$, it is easy to see that the algorithm works with very less number of qubits, even lesser than $\log (N)$, N being the number of data points in the dataset, have a polynomial  gate and operation complexity, and with measurements polynomial in qubit number. We can also see that with an assumed read-out error, it out performs its ideal case counter-part. This makes the algorithm extremely suitable for the Noisy Intermediate Scale Quantum (NISQ) computers. 

\section{Conclusion}\label{sec 5}
In this work, we have presented two NISQ friendly algorithms for unsupervised data clustering. QHCA has a time complexity of $O(KD_{max}N\log N)$ which is better than the classical hierarchical clustering algorithms,( $O(2^N)$ for divisive and $O(N^3)$ for agglomerative). The quantum part of both the algorithms is independent of number of datapoints. We have shown that QHCA provides better results than the classical divisive clustering on the linearly non-separable dataset, even with very low number of measurements. We further showed that for the Churritz cities dataset, the algorithm is able to cluster with only $7$ qubits. The second algorithm is a prepare and measure algorithm that makes use of unsharp measurements to cluster similar data points together. We show a successful classification of the Wisconsin cancer dataset with high accuracy with this algorithm, and further show that we need only polynomial measurements to perform this task. We also provide proof that under an assumption of measurement noise, the algorithm performs better than an ideal scenario. For this work, we have taken the width of the effect operator constant, and have performed exhaustive search to find the optimal measurement limit. As a future approach, we propose to optimize these parameters to achieve better results. Further, the realization of the algorithms on a real quantum device is beyond the scope of this paper, however, we look forward to examine the correlation between noises present on a real NISQ device and unsharp measurements, and investigate its effect on the accuracy of clustering in the future.

\bibliography{ref}


\begin{thebibliography}{34}
\ifx \bisbn   \undefined \def \bisbn  #1{ISBN #1}\fi
\ifx \binits  \undefined \def \binits#1{#1}\fi
\ifx \bauthor  \undefined \def \bauthor#1{#1}\fi
\ifx \batitle  \undefined \def \batitle#1{#1}\fi
\ifx \bjtitle  \undefined \def \bjtitle#1{#1}\fi
\ifx \bvolume  \undefined \def \bvolume#1{\textbf{#1}}\fi
\ifx \byear  \undefined \def \byear#1{#1}\fi
\ifx \bissue  \undefined \def \bissue#1{#1}\fi
\ifx \bfpage  \undefined \def \bfpage#1{#1}\fi
\ifx \blpage  \undefined \def \blpage #1{#1}\fi
\ifx \burl  \undefined \def \burl#1{\textsf{#1}}\fi
\ifx \doiurl  \undefined \def \doiurl#1{\url{https://doi.org/#1}}\fi
\ifx \betal  \undefined \def \betal{\textit{et al.}}\fi
\ifx \binstitute  \undefined \def \binstitute#1{#1}\fi
\ifx \binstitutionaled  \undefined \def \binstitutionaled#1{#1}\fi
\ifx \bctitle  \undefined \def \bctitle#1{#1}\fi
\ifx \beditor  \undefined \def \beditor#1{#1}\fi
\ifx \bpublisher  \undefined \def \bpublisher#1{#1}\fi
\ifx \bbtitle  \undefined \def \bbtitle#1{#1}\fi
\ifx \bedition  \undefined \def \bedition#1{#1}\fi
\ifx \bseriesno  \undefined \def \bseriesno#1{#1}\fi
\ifx \blocation  \undefined \def \blocation#1{#1}\fi
\ifx \bsertitle  \undefined \def \bsertitle#1{#1}\fi
\ifx \bsnm \undefined \def \bsnm#1{#1}\fi
\ifx \bsuffix \undefined \def \bsuffix#1{#1}\fi
\ifx \bparticle \undefined \def \bparticle#1{#1}\fi
\ifx \barticle \undefined \def \barticle#1{#1}\fi
\bibcommenthead
\ifx \bconfdate \undefined \def \bconfdate #1{#1}\fi
\ifx \botherref \undefined \def \botherref #1{#1}\fi
\ifx \url \undefined \def \url#1{\textsf{#1}}\fi
\ifx \bchapter \undefined \def \bchapter#1{#1}\fi
\ifx \bbook \undefined \def \bbook#1{#1}\fi
\ifx \bcomment \undefined \def \bcomment#1{#1}\fi
\ifx \oauthor \undefined \def \oauthor#1{#1}\fi
\ifx \citeauthoryear \undefined \def \citeauthoryear#1{#1}\fi
\ifx \endbibitem  \undefined \def \endbibitem {}\fi
\ifx \bconflocation  \undefined \def \bconflocation#1{#1}\fi
\ifx \arxivurl  \undefined \def \arxivurl#1{\textsf{#1}}\fi
\csname PreBibitemsHook\endcsname

\bibitem[\protect\citeauthoryear{Arute et~al.}{2019}]{arute2019quantum}
\begin{barticle}
\bauthor{\bsnm{Arute}, \binits{F.}},
\bauthor{\bsnm{Arya}, \binits{K.}},
\bauthor{\bsnm{Babbush}, \binits{R.}},
\bauthor{\bsnm{Bacon}, \binits{D.}},
\bauthor{\bsnm{Bardin}, \binits{J.C.}},
\bauthor{\bsnm{Barends}, \binits{R.}},
\bauthor{\bsnm{Biswas}, \binits{R.}},
\bauthor{\bsnm{Boixo}, \binits{S.}},
\bauthor{\bsnm{Brandao}, \binits{F.G.}},
\bauthor{\bsnm{Buell}, \binits{D.A.}}, \betal:
\batitle{Quantum supremacy using a programmable superconducting processor}.
\bjtitle{Nature}
\bvolume{574}(\bissue{7779}),
\bfpage{505}--\blpage{510}
(\byear{2019})
\doiurl{10.1038/s41586-019-1666-5}
\end{barticle}
\endbibitem

\bibitem[\protect\citeauthoryear{Feynman et~al.}{2018}]{feynman2018simulating}
\begin{botherref}
\oauthor{\bsnm{Feynman}, \binits{R.P.}}, et al.:
Simulating physics with computers.
Int. j. Theor. phys
\textbf{21}(6/7)
(2018)
\end{botherref}
\endbibitem

\bibitem[\protect\citeauthoryear{Deutsch and Jozsa}{1992}]{deutsch1992rapid}
\begin{barticle}
\bauthor{\bsnm{Deutsch}, \binits{D.}},
\bauthor{\bsnm{Jozsa}, \binits{R.}}:
\batitle{Rapid solution of problems by quantum computation}.
\bjtitle{Proc. Math. Phys. Eng. Sci.}
\bvolume{439}(\bissue{1907}),
\bfpage{553}--\blpage{558}
(\byear{1992})
\doiurl{10.1098/rspa.1992.0167}
\end{barticle}
\endbibitem

\bibitem[\protect\citeauthoryear{Collins et~al.}{1998}]{collins1998deutsch}
\begin{barticle}
\bauthor{\bsnm{Collins}, \binits{D.}},
\bauthor{\bsnm{Kim}, \binits{K.}},
\bauthor{\bsnm{Holton}, \binits{W.}}:
\batitle{Deutsch-jozsa algorithm as a test of quantum computation}.
\bjtitle{Phys. Rev. A}
\bvolume{58}(\bissue{3}),
\bfpage{1633}
(\byear{1998})
\doiurl{10.1103/PhysRevA.58.R1633}
\end{barticle}
\endbibitem

\bibitem[\protect\citeauthoryear{Bernstein and Vazirani}{1993}]{10.1145/167088.167097}
\begin{bchapter}
\bauthor{\bsnm{Bernstein}, \binits{E.}},
\bauthor{\bsnm{Vazirani}, \binits{U.}}:
\bctitle{Quantum complexity theory}.
In: \bbtitle{Proc. Annu. ACM Symp. Theory Comput.}
\bsertitle{STOC '93},
pp. \bfpage{11}--\blpage{20}.
\bpublisher{Association for Computing Machinery},
\blocation{New York, NY, USA}
(\byear{1993}).
\doiurl{10.1145/167088.167097} .
\burl{https://doi.org/10.1145/167088.167097}
\end{bchapter}
\endbibitem

\bibitem[\protect\citeauthoryear{Shor}{1997}]{10.1137/S0097539795293172}
\begin{barticle}
\bauthor{\bsnm{Shor}, \binits{P.W.}}:
\batitle{Polynomial-time algorithms for prime factorization and discrete logarithms on a quantum computer}.
\bjtitle{SIAM J. Comput.}
\bvolume{26}(\bissue{5}),
\bfpage{1484}--\blpage{1509}
(\byear{1997})
\doiurl{10.1137/S0097539795293172}
\end{barticle}
\endbibitem

\bibitem[\protect\citeauthoryear{Grover}{1996}]{10.1145/237814.237866}
\begin{bchapter}
\bauthor{\bsnm{Grover}, \binits{L.K.}}:
\bctitle{A fast quantum mechanical algorithm for database search}.
In: \bbtitle{Proc. Annu. ACM Symp. Theory Comput.}
\bsertitle{STOC '96},
pp. \bfpage{212}--\blpage{219}.
\bpublisher{Association for Computing Machinery},
\blocation{New York, NY, USA}
(\byear{1996}).
\doiurl{10.1145/237814.237866} .
\burl{https://doi.org/10.1145/237814.237866}
\end{bchapter}
\endbibitem

\bibitem[\protect\citeauthoryear{Harrow et~al.}{2009}]{PhysRevLett.103.150502}
\begin{barticle}
\bauthor{\bsnm{Harrow}, \binits{A.W.}},
\bauthor{\bsnm{Hassidim}, \binits{A.}},
\bauthor{\bsnm{Lloyd}, \binits{S.}}:
\batitle{Quantum algorithm for linear systems of equations}.
\bjtitle{Phys. Rev. Lett.}
\bvolume{103},
\bfpage{150502}
(\byear{2009})
\doiurl{10.1103/PhysRevLett.103.150502}
\end{barticle}
\endbibitem

\bibitem[\protect\citeauthoryear{Horn and Gottlieb}{2001}]{PhysRevLett.88.018702}
\begin{barticle}
\bauthor{\bsnm{Horn}, \binits{D.}},
\bauthor{\bsnm{Gottlieb}, \binits{A.}}:
\batitle{Algorithm for data clustering in pattern recognition problems based on quantum mechanics}.
\bjtitle{Phys. Rev. Lett.}
\bvolume{88},
\bfpage{018702}
(\byear{2001})
\doiurl{10.1103/PhysRevLett.88.018702}
\end{barticle}
\endbibitem

\bibitem[\protect\citeauthoryear{Diday and Simon}{1976}]{Diday1976}
\begin{bbook}
\bauthor{\bsnm{Diday}, \binits{E.}},
\bauthor{\bsnm{Simon}, \binits{J.C.}}:
In: \beditor{\bsnm{Fu}, \binits{K.S.}} (ed.)
\bbtitle{Clustering Analysis},
pp. \bfpage{47}--\blpage{94}.
\bpublisher{Springer},
\blocation{Berlin, Heidelberg}
(\byear{1976}).
\doiurl{10.1007/978-3-642-96303-2_3} .
\burl{https://doi.org/10.1007/978-3-642-96303-2_3}
\end{bbook}
\endbibitem

\bibitem[\protect\citeauthoryear{Mandal et~al.}{2021}]{9605278}
\begin{bchapter}
\bauthor{\bsnm{Mandal}, \binits{A.}},
\bauthor{\bsnm{Banerjee}, \binits{S.}},
\bauthor{\bsnm{Panigrahi}, \binits{P.K.}}:
\bctitle{Quantum image representation on clusters}.
In: \bbtitle{2021 IEEE International Conference on Quantum Computing and Engineering (QCE)},
pp. \bfpage{89}--\blpage{99}
(\byear{2021}).
\doiurl{10.1109/QCE52317.2021.00025}
\end{bchapter}
\endbibitem

\bibitem[\protect\citeauthoryear{Ahuja et~al.}{2020}]{ahuja2020}
\begin{bbook}
\bauthor{\bsnm{Ahuja}, \binits{R.}},
\bauthor{\bsnm{Chug}, \binits{A.}},
\bauthor{\bsnm{Gupta}, \binits{S.}},
\bauthor{\bsnm{Ahuja}, \binits{P.}},
\bauthor{\bsnm{Kohli}, \binits{S.}}:
In: \beditor{\bsnm{Yang}, \binits{X.-S.}},
\beditor{\bsnm{He}, \binits{X.-S.}} (eds.)
\bbtitle{Classification and Clustering Algorithms of Machine Learning with their Applications},
pp. \bfpage{225}--\blpage{248}.
\bpublisher{Springer},
\blocation{Cham}
(\byear{2020}).
\doiurl{10.1007/978-3-030-28553-1_11} .
\burl{https://doi.org/10.1007/978-3-030-28553-1_11}
\end{bbook}
\endbibitem

\bibitem[\protect\citeauthoryear{Everitt et~al.}{2011}]{everitt2011cluster}
\begin{botherref}
\oauthor{\bsnm{Everitt}, \binits{B.S.}},
\oauthor{\bsnm{Landau}, \binits{S.}},
\oauthor{\bsnm{Leese}, \binits{M.}},
\oauthor{\bsnm{Stahl}, \binits{D.}}:
Cluster analysis: Wiley series in probability and statistics.
Southern Gate, Chichester, West SussexUnited Kingdom: John Wiley \& Sons
(2011)
\end{botherref}
\endbibitem

\bibitem[\protect\citeauthoryear{Kaufman and Rousseeuw}{2009}]{kaufman2009finding}
\begin{bbook}
\bauthor{\bsnm{Kaufman}, \binits{L.}},
\bauthor{\bsnm{Rousseeuw}, \binits{P.J.}}:
\bbtitle{Finding Groups in Data: an Introduction to Cluster Analysis}.
\bpublisher{John Wiley \& Sons},
\blocation{United States}
(\byear{2009}).
\doiurl{10.1002/9780470316801}
\end{bbook}
\endbibitem

\bibitem[\protect\citeauthoryear{Hartigan and Wong}{1979}]{Hartigan1979KMeans}
\begin{barticle}
\bauthor{\bsnm{Hartigan}, \binits{J.A.}},
\bauthor{\bsnm{Wong}, \binits{M.A.}}:
\batitle{Algorithm {AS} 136: A {K-Means} clustering algorithm}.
\bjtitle{App. Stat.}
\bvolume{28}(\bissue{1}),
\bfpage{100}--\blpage{108}
(\byear{1979})
\doiurl{10.2307/2346830}
\end{barticle}
\endbibitem

\bibitem[\protect\citeauthoryear{Ester et~al.}{1996}]{10.5555/3001460.3001507}
\begin{bchapter}
\bauthor{\bsnm{Ester}, \binits{M.}},
\bauthor{\bsnm{Kriegel}, \binits{H.-P.}},
\bauthor{\bsnm{Sander}, \binits{J.}},
\bauthor{\bsnm{Xu}, \binits{X.}}:
\bctitle{A density-based algorithm for discovering clusters in large spatial databases with noise}.
In: \bbtitle{Proceedings of the Second International Conference on Knowledge Discovery and Data Mining}.
\bsertitle{KDD'96},
pp. \bfpage{226}--\blpage{231}.
\bpublisher{AAAI Press}, \blocation{???}
(\byear{1996}).
\burl{https://www.bibsonomy.org/bibtex/289e04610c1b5f2fa147398826b502a2d/nosebrain}
\end{bchapter}
\endbibitem

\bibitem[\protect\citeauthoryear{D{\"u}rr et~al.}{2006}]{durr2006quantum}
\begin{barticle}
\bauthor{\bsnm{D{\"u}rr}, \binits{C.}},
\bauthor{\bsnm{Heiligman}, \binits{M.}},
\bauthor{\bsnm{HOyer}, \binits{P.}},
\bauthor{\bsnm{Mhalla}, \binits{M.}}:
\batitle{Quantum query complexity of some graph problems}.
\bjtitle{SIAM Journal on Computing}
\bvolume{35}(\bissue{6}),
\bfpage{1310}--\blpage{1328}
(\byear{2006})
\end{barticle}
\endbibitem

\bibitem[\protect\citeauthoryear{Li et~al.}{2009}]{li2009novel}
\begin{barticle}
\bauthor{\bsnm{Li}, \binits{Q.}},
\bauthor{\bsnm{He}, \binits{Y.}},
\bauthor{\bsnm{Jiang}, \binits{J.-p.}}:
\batitle{A novel clustering algorithm based on quantum games}.
\bjtitle{J. Phys. A: Math. and Theor.}
\bvolume{42}(\bissue{44}),
\bfpage{445303}
(\byear{2009})
\end{barticle}
\endbibitem

\bibitem[\protect\citeauthoryear{Li et~al.}{2011}]{li2011hybrid}
\begin{barticle}
\bauthor{\bsnm{Li}, \binits{Q.}},
\bauthor{\bsnm{He}, \binits{Y.}},
\bauthor{\bsnm{Jiang}, \binits{J.-p.}}:
\batitle{A hybrid classical-quantum clustering algorithm based on quantum walks}.
\bjtitle{Quantum Inf. Process.}
\bvolume{10},
\bfpage{13}--\blpage{26}
(\byear{2011})
\end{barticle}
\endbibitem

\bibitem[\protect\citeauthoryear{Yu et~al.}{2010}]{yu2010quantum}
\begin{barticle}
\bauthor{\bsnm{Yu}, \binits{Y.}},
\bauthor{\bsnm{Qian}, \binits{F.}},
\bauthor{\bsnm{Liu}, \binits{H.}}:
\batitle{Quantum clustering-based weighted linear programming support vector regression for multivariable nonlinear problem}.
\bjtitle{Soft Comput.}
\bvolume{14}(\bissue{9}),
\bfpage{921}--\blpage{929}
(\byear{2010})
\end{barticle}
\endbibitem

\bibitem[\protect\citeauthoryear{A\"{\i}meur et~al.}{2007}]{10.1145/1273496.1273497}
\begin{bchapter}
\bauthor{\bsnm{A\"{\i}meur}, \binits{E.}},
\bauthor{\bsnm{Brassard}, \binits{G.}},
\bauthor{\bsnm{Gambs}, \binits{S.}}:
\bctitle{Quantum clustering algorithms}.
In: \bbtitle{Proceedings of the 24th International Conference on Machine Learning}.
\bsertitle{ICML '07},
pp. \bfpage{1}--\blpage{8}.
\bpublisher{Association for Computing Machinery},
\blocation{New York, NY, USA}
(\byear{2007}).
\doiurl{10.1145/1273496.1273497} .
\burl{https://doi.org/10.1145/1273496.1273497}
\end{bchapter}
\endbibitem

\bibitem[\protect\citeauthoryear{Casaña-Eslava et~al.}{2020}]{CASANAESLAVA2020105567}
\begin{barticle}
\bauthor{\bsnm{Casaña-Eslava}, \binits{R.V.}},
\bauthor{\bsnm{Lisboa}, \binits{P.J.G.}},
\bauthor{\bsnm{Ortega-Martorell}, \binits{S.}},
\bauthor{\bsnm{Jarman}, \binits{I.H.}},
\bauthor{\bsnm{Martín-Guerrero}, \binits{J.D.}}:
\batitle{Probabilistic quantum clustering}.
\bjtitle{Knowledge-Based Systems}
\bvolume{194},
\bfpage{105567}
(\byear{2020})
\doiurl{10.1016/j.knosys.2020.105567}
\end{barticle}
\endbibitem

\bibitem[\protect\citeauthoryear{Bermejo and Or{\'u}s}{2023}]{bermejo2023variational}
\begin{barticle}
\bauthor{\bsnm{Bermejo}, \binits{P.}},
\bauthor{\bsnm{Or{\'u}s}, \binits{R.}}:
\batitle{Variational quantum and quantum-inspired clustering}.
\bjtitle{Scientific Reports}
\bvolume{13}(\bissue{1}),
\bfpage{13284}
(\byear{2023})
\doiurl{10.1038/s41598-023-39771-6}
\end{barticle}
\endbibitem

\bibitem[\protect\citeauthoryear{Khan et~al.}{2019}]{Ahsan2019}
\begin{botherref}
\oauthor{\bsnm{Khan}, \binits{S.U.}},
\oauthor{\bsnm{Awan}, \binits{A.J.}},
\oauthor{\bsnm{Vall{-}Llosera}, \binits{G.}}:
K-means clustering on noisy intermediate scale quantum computers.
CoRR
\textbf{abs/1909.12183}
(2019)
{\href{https://arxiv.org/abs/1909.12183}{{1909.12183}}}
\end{botherref}
\endbibitem

\bibitem[\protect\citeauthoryear{Kavitha and Kaulgud}{2023}]{kavitha2023quantum}
\begin{barticle}
\bauthor{\bsnm{Kavitha}, \binits{S.}},
\bauthor{\bsnm{Kaulgud}, \binits{N.}}:
\batitle{Quantum k-means clustering method for detecting heart disease using quantum circuit approach}.
\bjtitle{Soft Computing}
\bvolume{27}(\bissue{18}),
\bfpage{13255}--\blpage{13268}
(\byear{2023})
\doiurl{10.1007/s00500-022-07200-x}
\end{barticle}
\endbibitem

\bibitem[\protect\citeauthoryear{Li et~al.}{2022}]{li2022quantum}
\begin{barticle}
\bauthor{\bsnm{Li}, \binits{Q.}},
\bauthor{\bsnm{Huang}, \binits{Y.}},
\bauthor{\bsnm{Jin}, \binits{S.}},
\bauthor{\bsnm{Hou}, \binits{X.}},
\bauthor{\bsnm{Wang}, \binits{X.}}:
\batitle{Quantum spectral clustering algorithm for unsupervised learning}.
\bjtitle{Science China Information Sciences}
\bvolume{65}(\bissue{10}),
\bfpage{200504}
(\byear{2022})
\doiurl{10.1007/s11432-022-3492-x}
\end{barticle}
\endbibitem

\bibitem[\protect\citeauthoryear{Gopalakrishnan et~al.}{2024}]{Luca2024}
\begin{botherref}
\oauthor{\bsnm{Gopalakrishnan}, \binits{D.}},
\oauthor{\bsnm{Dellantonio}, \binits{L.}},
\oauthor{\bsnm{Pilato}, \binits{A.D.}},
\oauthor{\bsnm{Redjeb}, \binits{W.}},
\oauthor{\bsnm{Pantaleo}, \binits{F.}},
\oauthor{\bsnm{Mosca}, \binits{M.}}:
qLUE: A Quantum Clustering Algorithm for Multi- Dimensional Datasets
(2024).
\url{https://arxiv.org/abs/2407.00357}
\end{botherref}
\endbibitem

\bibitem[\protect\citeauthoryear{Day and Edelsbrunner}{1984}]{day1984efficient}
\begin{barticle}
\bauthor{\bsnm{Day}, \binits{W.H.}},
\bauthor{\bsnm{Edelsbrunner}, \binits{H.}}:
\batitle{Efficient algorithms for agglomerative hierarchical clustering methods}.
\bjtitle{J. Classif.}
\bvolume{1}(\bissue{1}),
\bfpage{7}--\blpage{24}
(\byear{1984})
\end{barticle}
\endbibitem

\bibitem[\protect\citeauthoryear{A{\"i}meur et~al.}{2013}]{aimeur:hal-00736948}
\begin{barticle}
\bauthor{\bsnm{A{\"i}meur}, \binits{E.}},
\bauthor{\bsnm{Brassard}, \binits{G.}},
\bauthor{\bsnm{Gambs}, \binits{S.}}:
\batitle{{Quantum speed-up for unsupervised learning}}.
\bjtitle{{Mach. Learn.}}
\bvolume{90}(\bissue{2}),
\bfpage{261}--\blpage{287}
(\byear{2013})
\doiurl{10.1007/s10994-012-5316-5}
\end{barticle}
\endbibitem

\bibitem[\protect\citeauthoryear{Foulis and Bennett}{1994}]{foulis1994effect}
\begin{barticle}
\bauthor{\bsnm{Foulis}, \binits{D.J.}},
\bauthor{\bsnm{Bennett}, \binits{M.K.}}:
\batitle{Effect algebras and unsharp quantum logics}.
\bjtitle{Found. Phys.}
\bvolume{24}(\bissue{10}),
\bfpage{1331}--\blpage{1352}
(\byear{1994})
\end{barticle}
\endbibitem

\bibitem[\protect\citeauthoryear{Shukla and Vedula}{2024}]{Shukla2024}
\begin{barticle}
\bauthor{\bsnm{Shukla}, \binits{A.}},
\bauthor{\bsnm{Vedula}, \binits{P.}}:
\batitle{An efficient quantum algorithm for preparation of uniform quantum superposition states}.
\bjtitle{Quantum Information Processing}
\bvolume{23}(\bissue{2}),
\bfpage{38}
(\byear{2024})
\doiurl{10.1007/s11128-024-04258-4}
\end{barticle}
\endbibitem

\bibitem[\protect\citeauthoryear{Mozafari et~al.}{2021}]{9506863}
\begin{barticle}
\bauthor{\bsnm{Mozafari}, \binits{F.}},
\bauthor{\bsnm{Riener}, \binits{H.}},
\bauthor{\bsnm{Soeken}, \binits{M.}},
\bauthor{\bsnm{De~Micheli}, \binits{G.}}:
\batitle{Efficient boolean methods for preparing uniform quantum states}.
\bjtitle{IEEE Transactions on Quantum Engineering}
\bvolume{2},
\bfpage{1}--\blpage{12}
(\byear{2021})
\doiurl{10.1109/TQE.2021.3101663}
\end{barticle}
\endbibitem

\bibitem[\protect\citeauthoryear{Barui et~al.}{2024}]{Barui2024}
\begin{barticle}
\bauthor{\bsnm{Barui}, \binits{A.}},
\bauthor{\bsnm{Pal}, \binits{M.}},
\bauthor{\bsnm{Panigrahi}, \binits{P.K.}}:
\batitle{A novel approach to threshold quantum images by using unsharp measurements}.
\bjtitle{Quantum Information Processing}
\bvolume{23}(\bissue{3}),
\bfpage{76}
(\byear{2024})
\doiurl{10.1007/s11128-024-04282-4}
\end{barticle}
\endbibitem

\bibitem[\protect\citeauthoryear{Reinelt}{2014}]{reinhelt2014tsplib}
\begin{botherref}
\oauthor{\bsnm{Reinelt}, \binits{G.}}:
$\{$TSPLIB$\}$: a library of sample instances for the TSP (and related problems) from various sources and of various types.
\url{http://comopt.ifi.uni-heidelberg.de/software/TSPLIB95/}
(2014)
\end{botherref}
\endbibitem

\end{thebibliography}

\end{document}